\documentclass[preprint,  review, 12pt]{elsarticle}
    
\usepackage{amssymb}
\usepackage{amsmath}
\usepackage{soul}
\usepackage[dvipsnames]{xcolor}
\usepackage[left]{lineno}
\usepackage{float}

\journal{Journal of Molecular Liquids}

\begin{document}

\begin{frontmatter}



\title{A molecular perspective on the emergence of long-range polar order from an isotropic fluid}


\author[inst1]{Aitor Erkoreka}

\affiliation[inst1]{organization={Department of Physics, Faculty of Science and Technology, University of the Basque Country UPV/EHU},
            city={Bilbao},
            country={Spain}}
            
\author[inst2]{Nerea Sebastián}
\author[inst2]{Alenka Mertelj}

\affiliation[inst2]{organization={Jožef Stefan Institute},
            city={Ljubljana},
            country={Slovenia}}

\author[inst1]{Josu Martinez-Perdiguero\corref{cor1}}
\ead{jesus.martinez@ehu.eus}
\cortext[cor1]{Corresponding author}

\begin{abstract}
The ferroelectric nematic phase (N$_{\text{F}}$) has quickly become the most studied system in liquid crystal research. In this work, we investigate the origin of such polar structure by studying a compound for which the N$_{\text{F}}$ phase directly follows the isotropic liquid phase on cooling, making it a particularly interesting system. Our experimental results evidence the presence of polar correlations already in the high-temperature phase, in which ferroelectric order can be induced under a sufficiently strong electric field. In the N$_{\text{F}}$ phase, molecular dynamics and polar correlations are investigated through detailed dynamic dielectric measurements, while second harmonic generation experiments evidence a large value of the main coefficient of the second order dielectric susceptibility tensor. Lastly, experimentally determined parameters are employed for calculations based on a recently proposed theoretical model for the stability of the N$_{\text{F}}$ phase. The obtained results suggest that the parallel alignment of dipoles is driven by a subtle interplay between electrostatic and excluded volume interactions.

\end{abstract}


\end{frontmatter}

\section{Introduction}
\label{sec:introduction}

Ferroelectricity is still one of the most dynamic fields of study in condensed matter. In fact, the least ordered system that can exhibit such a property, namely the ferroelectric nematic liquid crystalline phase, was only recently discovered \cite{mandle_nematic_2017, nishikawa_fluid_2017, chen_first-principles_2020, lavrentovich_review, nf_review}. As opposed to the conventional nematic phase (N), the ferroelectric nematic phase (N$_{\text{F}}$) exhibits spontaneous polar order of dipole moments, which breaks the inversion symmetry of the director $\mathbf{n}$, i.e. the unit vector that specifies the average direction of molecular orientation. Materials exhibiting this phase present spontaneous polarization values of the order of $\mu$C/cm$^2$ \cite{nishikawa_fluid_2017, chen_first-principles_2020, brown_multiple_2021}, unprecedented in soft matter, and large nonlinear optical coefficients \cite{folcia_ferroelectric_2022, li_how_2021, Nishikawa_giant}.

Along with the discovery of the N$_{\text{F}}$ phase other novel polar phases have been discovered, including a splay-modulated antiferroelectric phase, termed as N$_{\text{S}}$ or SmZ$_{\text{A}}$ \cite{ferroelastic, chen_smectic_2023}, and an orthogonal smectic phase, called SmA$_{\text{F}}$, in which the polarization lies along the layer normal \cite{chen_smectic_a}. The richness of these systems, coupled with the fact that their origin is still poorly understood, has made them a hot topic, not only in liquid crystal (LC) research, but in condensed matter research in general. In addition, the applicability of these materials in future devices should not be forgotten, and constitutes an area of research on its own \cite{nf_review, ferroelectric_applications}.

A general feature of ferroelectric nemagotens is their large dipole moment \cite{Nishikawa_giant}. To which extent the electrostatic, dipolar or excluded volume interactions facilitate the parallel alignment of dipoles is still a matter of debate. There are numerous studies, many of which were carried out before the actual discovery of the N$_{\text{F}}$ phase, that focus on either the electric dipole strength or the non-centrosymmetric shape of the molecules \cite{palffy-muhoray_ferroelectric_1988, biscarini_head-tail_nodate, wei_orientational_1992, lee_ferroelectric_1994, berardi_ferroelectric_2001}. Both experimental and theoretical approaches are needed to shed light on this issue. On the experimental side, broadband dielectric spectroscopy (BDS) has proven to be a valuable tool in order to study the molecular dynamics of these systems. In fact, some previous studies have shown complex relaxation modes in the prototypical ferroelectric nematogens RM734 and DIO \cite{brown_multiple_2021, yadav_polar_2022, erkoreka_rm734, erkoreka_dio}, which were recently interpreted in terms of underlying collective and non-collective molecular processes \cite{erkoreka_dio}. Furthermore, the extremely large values of the dielectric permittivity in the N$_{\text{F}}$ phase (between $10^3$ and $10^5$ \cite{brown_multiple_2021, yadav_polar_2022, erkoreka_rm734, erkoreka_dio}) were attributed to spurious effects caused by the confinement of the material inside a measurement cell \cite{erkoreka_dio}. On the theoretical and computational sides, studies ranging from Monte Carlo to fully atomistic molecular dynamics (MD) simulations can offer interesting insights \cite{chen_first-principles_2020,dft_ns,mandle_molecular_2021, dft_vs_mc_2023, tapered_2023, manabe_md}. Moreover, the development of simplified theoretical models should also be sought after, since they offer the possibility of reducing a complex system into the fundamental physical principles that govern its behavior. One example of such a model was recently developed by Madhusudana \cite{madhusudana}. He proposed that the electron distribution of most ferroelectric nematogens can be described by longitudinal surface charge density waves. Calculating the electrostatic interaction energy between a set of molecules with such a charge distribution, he was able to predict the stability of the N$_{\text{F}}$ phase.

In this work, we study a material, hereinafter UUQU-4-N, which exhibits an uncommon direct isotropic (Iso) to N$_{\text{F}}$ transition and is the first compound reported to exhibit the N$_{\text{F}}$ phase close to room temperature \cite{manabe_ferroelectric_2021,manabe_eremin}. It is therefore an interesting system to gain insights into the emergence of polar order. To this end, we have performed X-ray diffraction (XRD), BDS and second harmonic generation (SHG) measurements. With such input parameters and to complete the picture, we have additionally investigated the role of intermolecular electrostatic interactions in this system based on quantum chemical calculations and an adapted version of Madhusudana's model \cite{madhusudana}.

\section{Materials and Methods}
\label{sec:methods}

\subsection{Material}
The studied ferroelectric nematogen is 4-((4$^{\prime}$-butyl-2$^{\prime}$,3,5,6$^{\prime}$-tetrafluoro-[1,1$^{\prime}$-biphenyl]-4-yl)difluoromethoxy)-2,6-difluorobenzonitrile, or UUQU-4-N. Details about the synthesis of this material can be found elsewhere \cite{manabe_original}. It exhibits a direct Iso-N$_{\text{F}}$ transition at $19.6^{\circ}$C on cooling. 

\subsection{X-ray diffraction measurements}
X-ray diffractograms were obtained using a Stoe Stadivari goniometer equipped with a Genix3D microfocus generator (Xenocs) and a Dectris Pilatus 100K detector. Monochromatic Cu K$_\alpha$ radiation ($\lambda=1.5406$ \AA{}) was used. The exposure time was $1$ minute. The temperature was varied using a nitrogen-gas Cryostream controller (Oxford Cryosystems) allowing for a temperature control within $0.1^{\circ}$C. The material was introduced in the Iso phase into a Lindemann capillary $0.6$ mm in diameter.

\subsection{Broadband dielectric spectroscopy}
The dynamic dielectric function $\varepsilon^*(f)=\varepsilon^{\prime}(f)-i\varepsilon^{\prime \prime}(f)$ was measured in the spectral range $f=$ 10 Hz--110 MHz. Two impedance analyzers were used for this purpose. Below 1 MHz, an Alpha-A impedance analyzer from Novocontrol Technologies GmbH was used. The high-frequency end was covered by an HP 4294A impedance analyzer. The sample was introduced between two circular gold-coated brass electrodes 5 mm in diameter forming a parallel-plate capacitor, the separation being fixed at $d=30$ $\mu$m by spherical silica spacers. This cell was then placed at the end of a modified HP 16091A coaxial test fixture, using a Quatro Cryostat for temperature control. No polyimide layers were used on the electrodes because, although they can be useful to obtain a proper alignment of LC molecules, they act as an additional large capacitance in series with the measurement cell and can lead to undesired charge accumulation, among other effects \cite{brown_multiple_2021, vaupotic_dielectric_2023, clark_dielectric_2022}. Experiments without DC bias were carried out with an oscillator voltage of $0.03$ V$_{\text{rms}}$. For DC bias measurements, however, the oscillator voltage was increased up to $1$ V$_{\text{rms}}$ so as to obtain a better signal-to-noise ratio. In all cases the sample was first heated up to $50^{\circ}$C and stabilized for 5 min at this temperature. The cooling rate was then set to $1^{\circ}$C/min down to $30^{\circ}$C and to $0.25^{\circ}$C/min thereafter. The stray capacitance of the measurement circuit was subtracted from the measured capacitance, and the complex dielectric permittivity was obtained by dividing this value by the capacitance of the empty cell. Finally, the experimental data were fitted to Havriliak-Negami (HN) relaxations with a conductivity term:

\begin{equation}
    \varepsilon^*(f) = \sum_{k} \frac{\Delta \varepsilon_k}{\left[1+\left(i \frac{f}{f_k}\right)^{\alpha_k} \right]^{\beta_k}} + \varepsilon_{\infty} + \frac{\sigma}{\varepsilon_0(i\,2\pi f)^{\lambda}}\mathrm{,}\label{HN_eq}
\end{equation}

\noindent
where $\Delta\varepsilon_k$, $f_k$, $\alpha_k$ and $\beta_k$ are respectively the dielectric strength, relaxation frequency and broadness exponents of mode $k$, $\varepsilon_{\infty}$ is the high-frequency dielectric permittivity, $\sigma$ is a measure of the conductivity, and $\lambda$ is an exponent between 0 and 1.

\subsection{Second harmonic generation measurements}
For SHG measurements, a Q-switched pulsed Nd:YAG laser operating at $\lambda=1064$ nm was employed. The beam was collimated, linearly polarized and the spot size was limited using a diaphragm. The intensity of the emitted second harmonic light ($\lambda=532$ nm) was measured by a photomultiplier after going through an analyzer. 

\subsection{Quantum chemical calculations}
Density functional theory (DFT) calculations were carried out using the ORCA 5.0.4 software package \cite{orca}. The molecular geometry was optimized and physical properties were calculated at the B3LYP/6-31G(d) level of theory. The obtained vibrational frequencies were checked in order to confirm that the final structure corresponded to an energy minimum. The hyperpolarizability tensor was calculated with Gaussian 16 \cite{g16} within coupled perturbed Hartree--Fock (CPHF) theory. Visualizations of the results were produced with Jmol \cite{jmol}.

\section{Results and discussion}
At the beginning of this section, a set of measurements on the mesomorphic and dielectric properties will be presented. These will not only serve to identify the different phases, but to determine some key quantities and gain some insights that will be relevant for the subsequent discussion on the formation of the N$_{\text{F}}$ phase. 

\begin{figure}[h!]
\begin{center}
\includegraphics[width=1.0\textwidth]{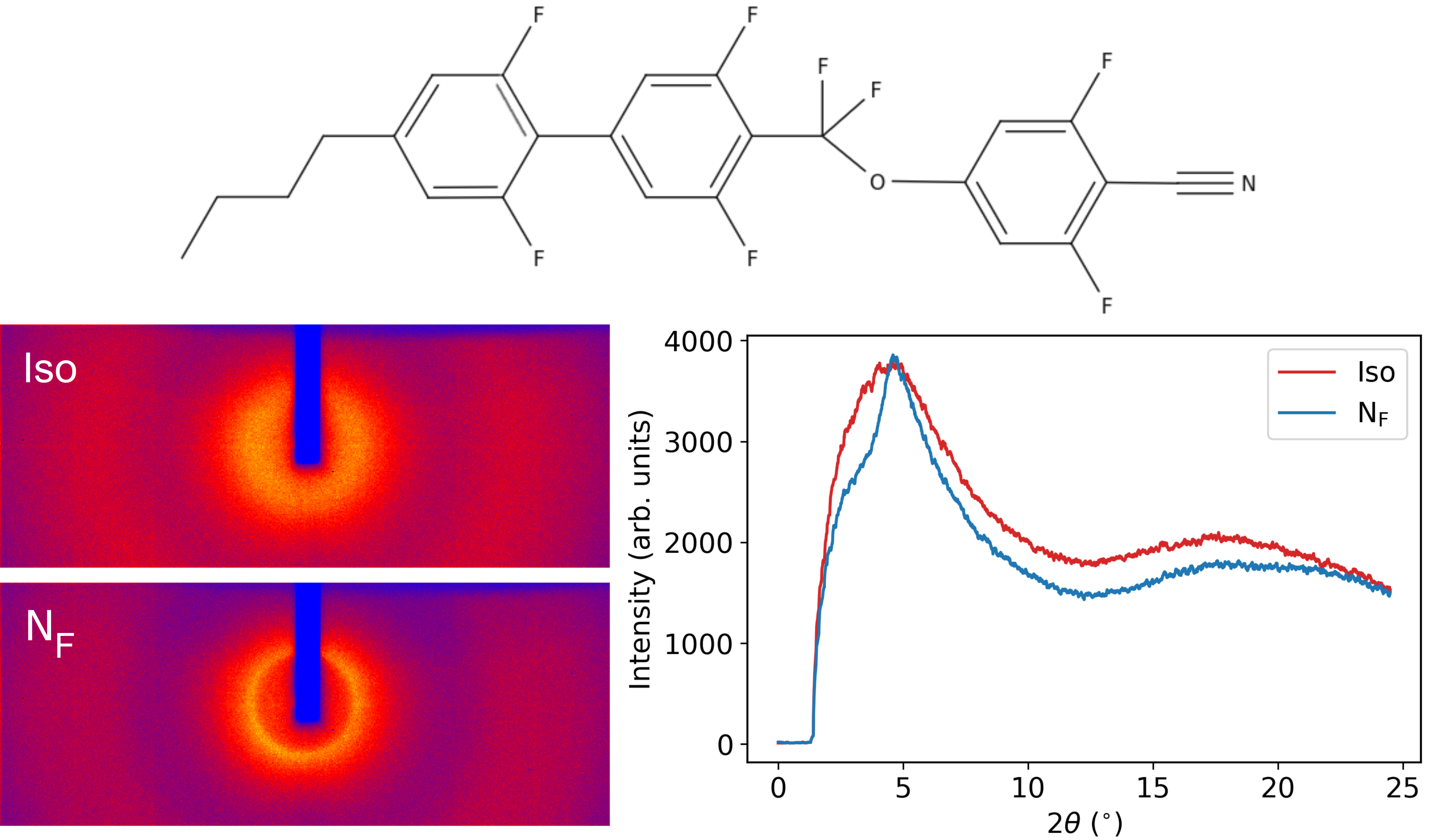}
\caption{\label{fig:xrd} Top: chemical structure of the UUQU-4-N molecule. Left: 2D X-ray diffractograms of UUQU-4-N in the Iso phase ($44^{\circ}$C) and N$_{\text{F}}$ phase ($16^{\circ}$C). Right: the integrated 1D pattern.}
\end{center}
\end{figure}

First of all, we determined the average intermolecular separations in the XRD patterns of the UUQU-4-N compound, as shown in Fig. \ref{fig:xrd}. The Iso phase is characterized by two broad diffraction peaks at $2\theta \approx 4.4^{\circ}$ and $17.6^{\circ}$, the corresponding $d$-spacings being $20$ \AA{} and $5$ \AA{}. Upon the transition to the N$_{\text{F}}$ phase, the small-angle peak becomes sharper while the diffuse halo moves to wider angles, confirming the lack of long-range positional order. The positions of these peaks are $4.6^{\circ}$ and $19.3^{\circ}$, corresponding to $19.2$ \AA{} and $4.6$ \AA{}, which are associated with the molecular length (compatible with that obtained from the DFT-optimized structure, see Table S1 in the Supplementary Material) and the $\pi$--$\pi$ stacking of aromatic cores, respectively. The sample then crystallized at $\sim 1^{\circ}$C (its XRD pattern can be found in Fig. S1).

\begin{figure}[h!]
\begin{center}
\includegraphics[width=0.7\textwidth]{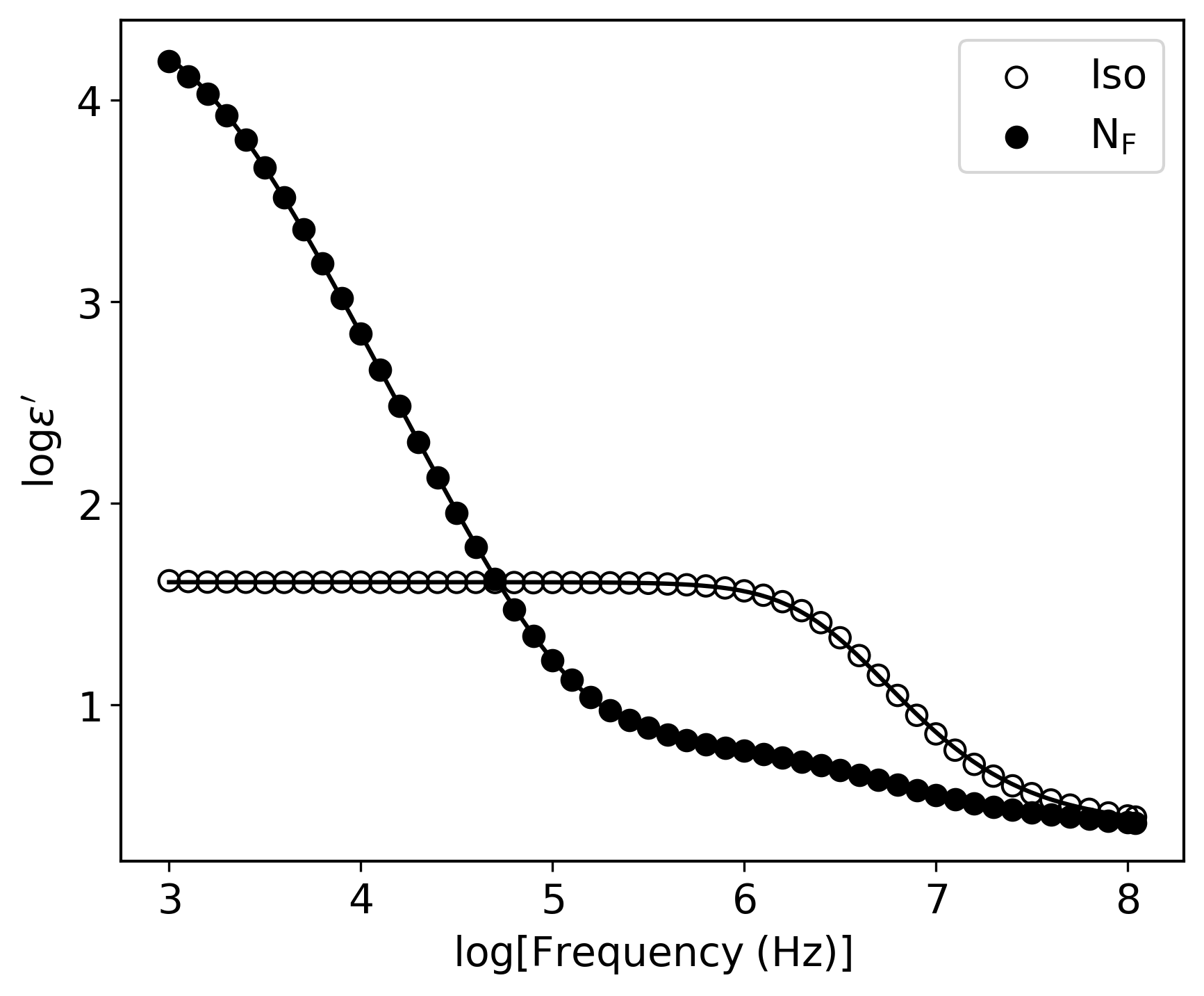}
\caption{\label{fig:spectra_at_2} Dielectric dispersions in the Iso ($45^{\circ}$C) and N$_{\text{F}}$ ($18^{\circ}$C) phases of UUQU-4-N measured in a 30 $\mu$m-thick cell with their corresponding fits.}
\end{center}
\end{figure}

In order to explore the molecular dynamics of the Iso--N$_{\text{F}}$ transition, we resorted to BDS. The spectrum of the real component of the permittivity at two temperatures corresponding to the Iso and N$_{\text{F}}$ phases can be seen in Fig. \ref{fig:spectra_at_2}. At first glance, it is evident that the Iso phase presents only one relaxation process, while two modes can be discerned in the N$_{\text{F}}$ phase. Especially noticeable is also the large increase in the permittivity values after the transition to the ferroelectric phase, reaching $\varepsilon' \sim 10^4$, which does not reflect an actual material property, as noted in other materials exhibiting the N$_{\text{F}}$ phase in cells of varying thickness for which $\varepsilon' \propto d$ was measured at low frequencies \cite{erkoreka_rm734, erkoreka_dio}.

\begin{figure}[h!]
\begin{center}
\includegraphics[width=0.6\textwidth]{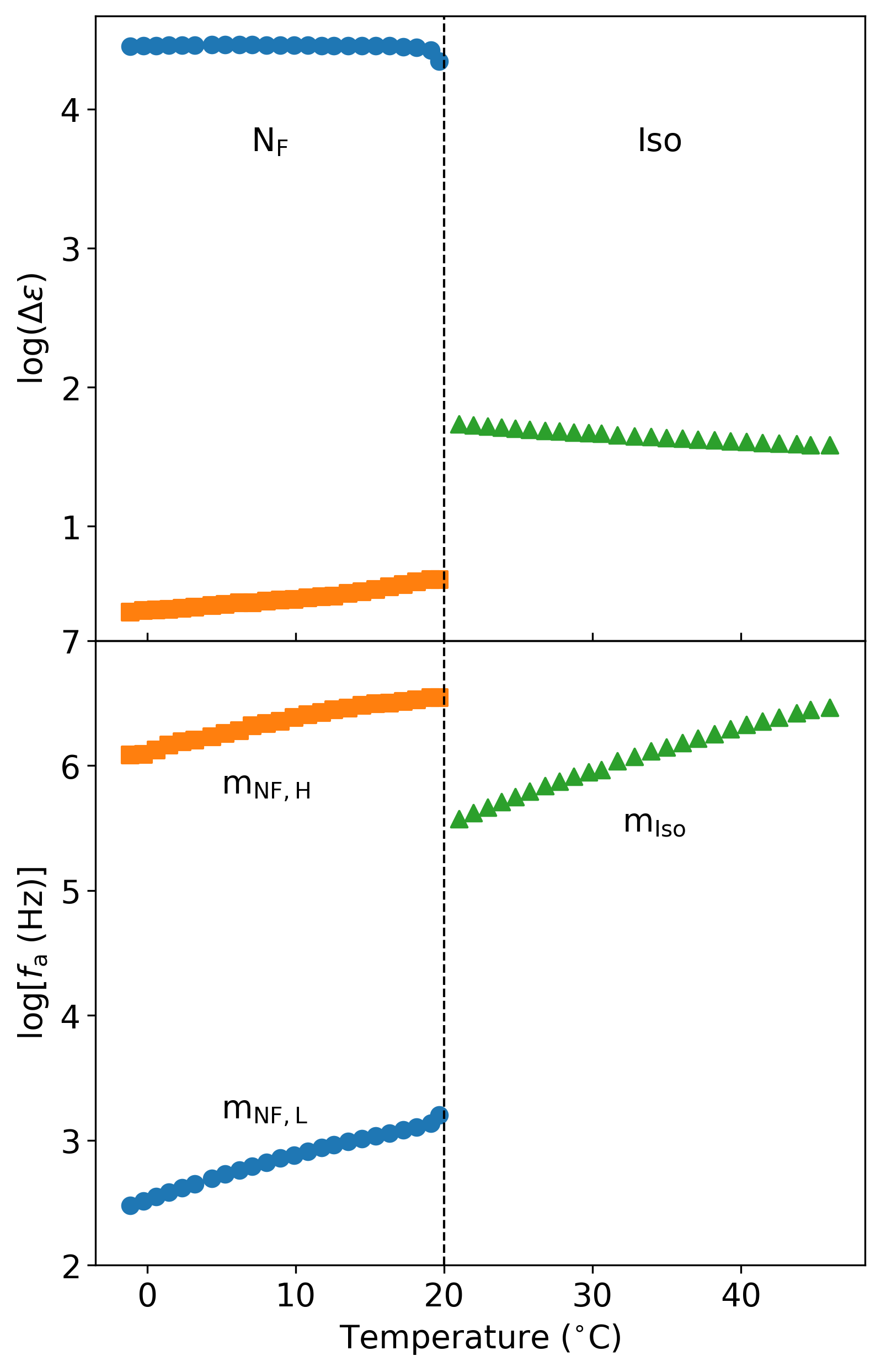}
\caption{\label{fig:4294-fit} Dielectric strengths ($\Delta\varepsilon$) and frequencies of maximum absorption ($f_a$) of the observed modes as a function of temperature obtained from fits to the HN formula (Equation \ref{HN_eq}).}
\end{center}
\end{figure}

Here we are interested in the evolution of polar correlations in the Iso phase leading to the formation of the N$_{\text{F}}$ phase. Fig. \ref{fig:4294-fit} shows the dielectric strengths and frequencies of maximum absorption obtained by fitting the observed relaxation processes to Eq. (\ref{HN_eq}) (fit examples and parameters can be found in Fig. S2 and Table S2). In the temperature interval corresponding to the Iso phase, only one relaxation process is observed, namely m$_{\text{Iso}}$. This mode can be to a great extent associated with the rotation of individual molecules around their short axis. It is worth noticing that the amplitude of this mode, which increases only slightly as the temperature is lowered, is quite large: $\Delta \varepsilon \sim 40$--$50$. As a comparison, the amplitude of the mode present in the Iso phase of 7CB, an ordinary nematic LC, is $\sim 7$ \cite{kremer_broadband_2003}. This can be partly attributed to the large dipole moment of the UUQU-4-N molecule of $\mu \sim 11$ D (see Table S1), whereas that of 7CB is $\sim 5$ D \cite{7cb_dipole}. Also, the absorption frequency of m$_{\text{Iso}}$ is unusually low, of the order of $1$ MHz, a fact that could be related to higher viscosities at the low N$_\text{F}$ temperatures of this material. Going back to the analogy with 7CB, the relaxation frequency of the corresponding mode is approximately $100$ MHz \cite{kremer_broadband_2003}. In addition, the absorption frequency of m$_{\text{Iso}}$ deviates slightly from the Arrhenius law and follows the Vogel-Fulcher-Tammann equation: $f_{\text{a}}=f_{\infty}\exp{\left[-A/(T-T_0)\right]}$, where $f_{\infty}$ and $A$ are constants and $T_0$ is the Vogel temperature (see Fig. S3). Deviation from Arrhenius behavior is typical of paraelectric to ferroelectric phase transitions, being related to the onset of cooperativity. The fact that the deviation is small and the growth of $\Delta \varepsilon$ contained indicates that, despite the direct transition, polar correlations are weak in the Iso phase.

Upon the transition to the N$_{\text{F}}$ phase, two modes can be distinguished, the director $\mathbf{n}$ and spontaneous polarization vector $\mathbf{P}$ lying parallel to the electrodes. The low-frequency mode, m$_{\text{NF,L}}$, is the Goldstone mode that is expected to appear in the ferroelectric phase due to spontaneous symmetry breaking \cite{vaupotic_dielectric_2023, clark_dielectric_2022, kremer_broadband_2003, haase_relaxation_2003}. As already discussed in previous articles \cite{erkoreka_rm734, erkoreka_dio}, its large amplitude can be attributed to the reorientation of the polarization which causes changes of surface bound charge $\mathbf{P}\cdot \mathbf{u}$ at the electrodes, $\mathbf{u}$ being the unit vector normal to the electrode surface \cite{Jackson}. Clark et al. refer to this phenomenon as PCG mode \cite{clark_dielectric_2022}. On the other hand, the fact that the absorption frequency of m$_{\text{NF,L}}$ decreases with decreasing temperature is just the result of an increasing material viscosity. The nature of m$_{\text{NF,H}}$ is not so clear. A continuous phenomenological model of the N$_{\text{F}}$ phase developed by Vauptič et al. \cite{vaupotic_dielectric_2023} predicts a collective high-frequency phason mode, also called optic mode \cite{erkoreka_dio}, which involves fluctations of $\mathbf{n}$ and $\mathbf{P}$ in counter-phase. For the prototypical ferroelectric nematogen DIO, we attributed the fast process of the N$_{\text{F}}$ phase, appearing at $\sim 10$--$20$ MHz, to the optic mode \cite{erkoreka_dio}. In the case of UUQU-4-N, m$_{\text{NF,H}}$ exhibits a similar amplitude but appears almost a decade below in frequency, which could in principle be explained by a higher viscosity. However, the dipole angle (i.e. the angle between the molecular dipole and the molecular long axis) of UUQU-4-N is virtually zero (see Table S1), in which case an optic mode would not be observable. In any case, it should be noted that DFT calculations are performed for a single molecule in a vacuum, whereas in a liquid the molecules can exhibit different conformations, resulting in changes in the direction of the dipole moment, among others. It would be interesting to study this issue through MD simulations, but it is beyond the scope of this paper.

\begin{figure}[h!]
\begin{center}
\includegraphics[width=0.7\textwidth]{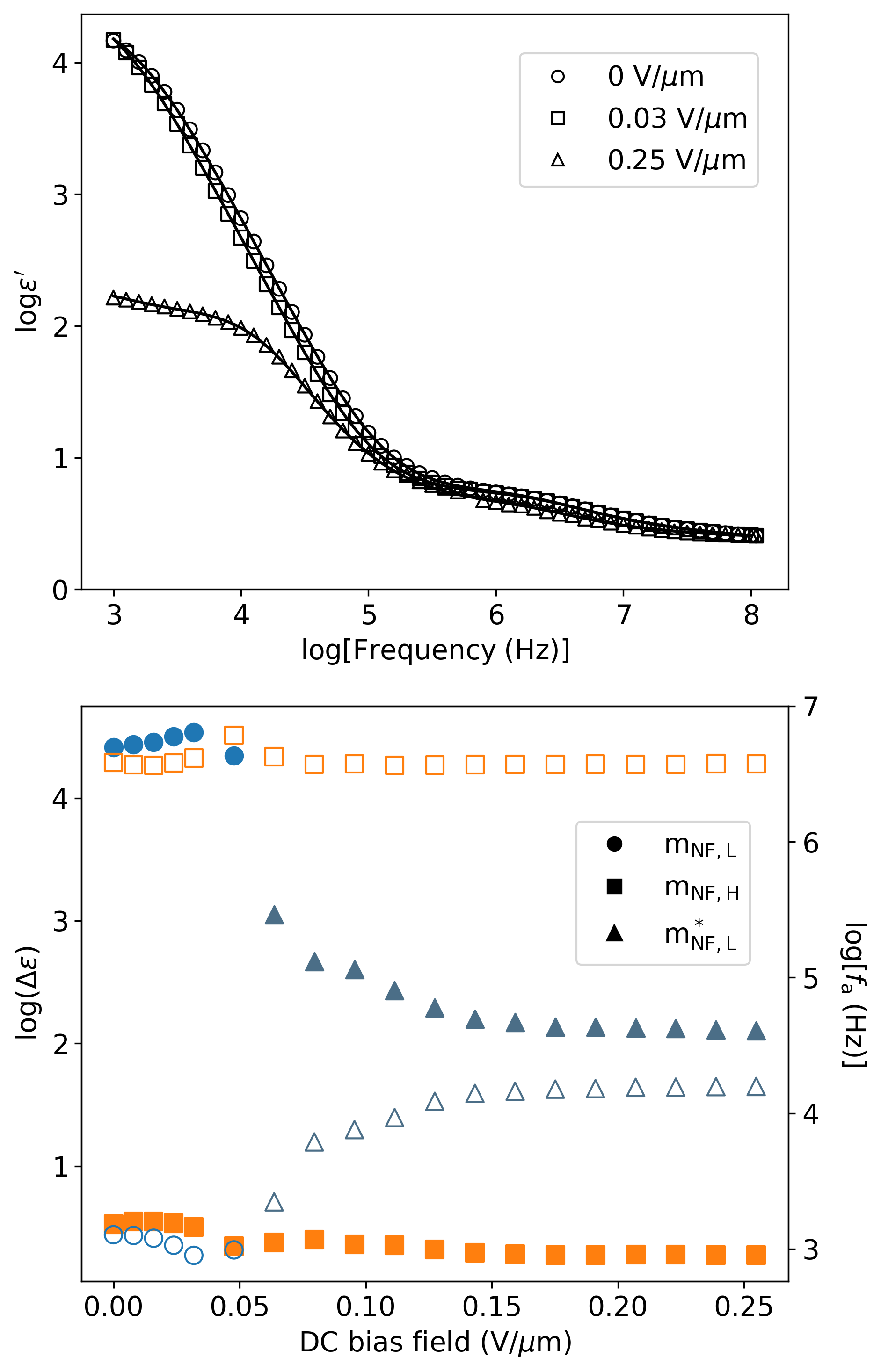}
\caption{\label{fig:dc_nf} Effect of a DC bias on the dielectric spectrum of the N$_{\text{F}}$ phase at $18^{\circ}$C. The permittivity spectra are shown at the top, while the fitted dielectric strengths (solid symbols) and absorption frequencies (empty symbols) are shown at the bottom.}
\end{center}
\end{figure}

In the following, we analyze how the dielectric spectra of both N$_{\text{F}}$ and Iso phases change upon applying a bias electric field. Fig. \ref{fig:dc_nf} shows the results at $18^{\circ}$C in the N$_{\text{F}}$ phase. As reported in a previous study \cite{erkoreka_dio}, at low DC bias fields the amplitude of m$_{\text{NF,L}}$ (Goldstone mode) slightly increases. At higher fields, however, the Goldstone mode is suppressed and the resulting mode m$^*_{\text{NF,L}}$ is the soft mode, which was recently interpreted as a collective flip-flop motion of molecules \cite{erkoreka_dio}. It is worth noticing that the fields required to observe this behavior are much smaller compared to our previous work on RM734 and DIO \cite{erkoreka_rm734, erkoreka_dio}. Nonetheless, the most striking and insightful feature of our data is that the measured permittivity values of all the studied ferroelectric nematic compounds (RM734, DIO and UUQU-4-N), once polarization reorientation processes have been suppressed, are almost the same ($\sim 150$) and independent of the sample thickness (as any intrinsic property should be). This is, in fact, not surprising considering that $\mu$ is very similar for all three materials. Therefore, we infer that measurements performed in the N$_{\text{F}}$ phase under a DC bias larger than the threshold value provide their intrinsic dielectric spectra, and we can estimate that $\varepsilon_{\parallel}\approx 150$. Regarding m$_{\text{NF,H}}$, its absorption frequency remains practically constant while its amplitude slightly decreases.

\begin{figure}[h!]
\begin{center}
\includegraphics[width=1.0\textwidth]{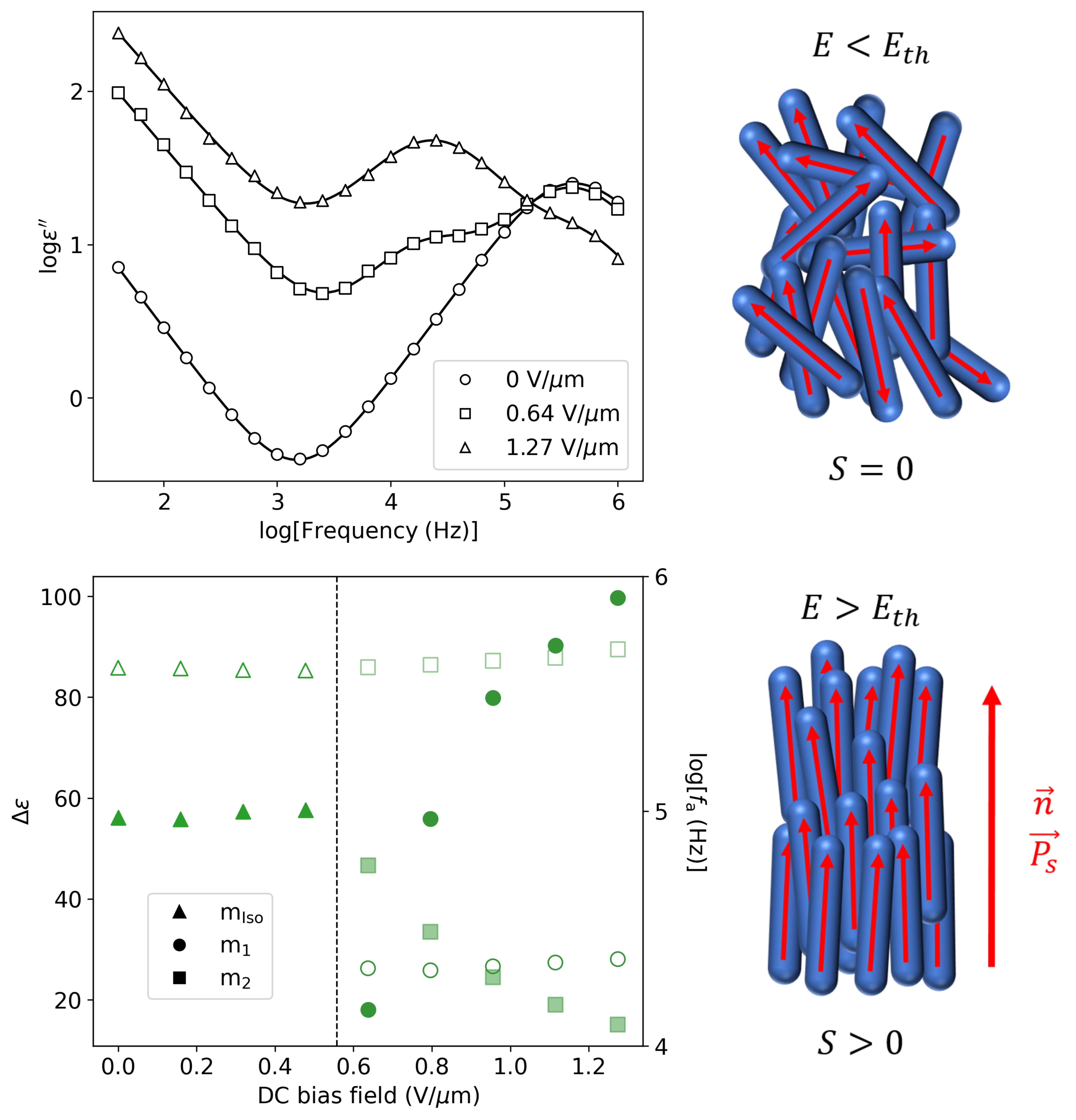}
\caption{\label{fig:dc_iso} Left: Effect of a DC bias on the dielectric spectrum of the Iso phase at $22^{\circ}$C. The dielectric loss spectra are shown at the top, while the fitted dielectric strengths (solid symbols) and absorption frequencies (empty symbols) are shown at the bottom. Right: Sketch of the effect of a DC bias at the molecular level. Below a threshold field $E_{\text{th}}$, the material remains in the Iso state. Above $E_{\text{th}}$, however, the N$_{\text{F}}$ phase is induced.}
\end{center}
\end{figure}

\begin{figure}[h!]
\begin{center}
\includegraphics[width=0.65\textwidth]{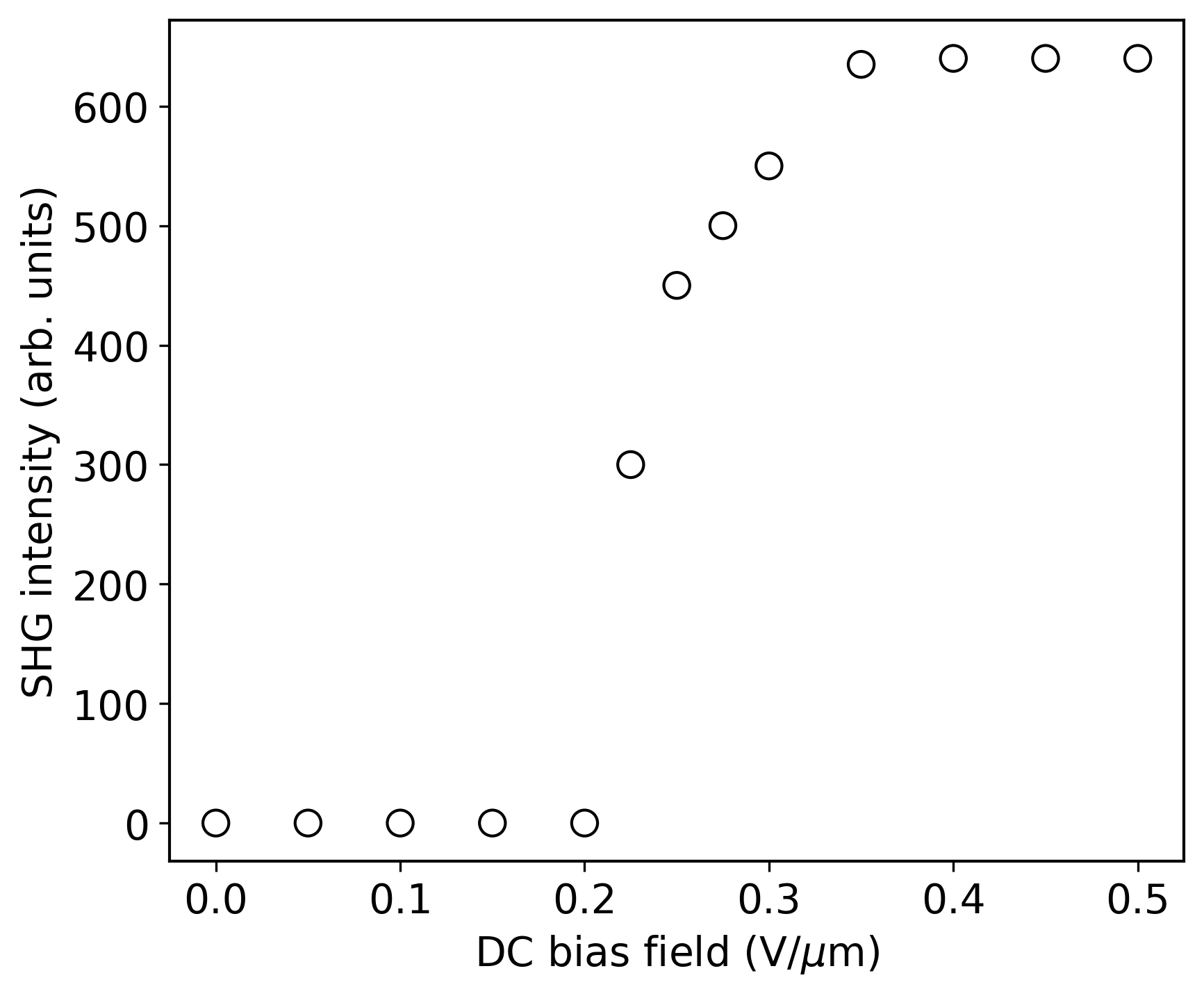}
\caption{\label{fig:shg} Measured SHG intensity as a function of the externally applied DC electric field at $22^{\circ}$C in the Iso phase.}
\end{center}
\end{figure}

Turning to the Iso phase, it is well established that a strong electric field can induce a N phase in ordinary LCs \cite{induced_nematic}. Since UUQU-4-N lacks a non-polar N phase and favors a parallel alignment of dipoles below the Iso phase, one might expect to induce a N$_{\text{F}}$ phase with a strong enough field. In fact, this phenomenon has recently been observed in a compound exhibiting a Iso--N$^{*}$--N$_{\text{F}}$ phase sequence, the N$^{*}$ (cholesteric) range being very narrow ($\sim 1^{\circ}$C) \cite{nf_critical}. The authors also found a critical point, analogous to a liquid-gas critical point, above which an isotropic phase continouously evolves into a N$_{\text{F}}$ phase under an increasing electric field. Our results $2^{\circ}$C above the transition shown in Fig. \ref{fig:dc_iso}, are compatible with their findings. Below a threshold electric field $E_{\text{th}}$, the sample remains in the Iso state and the dielectric spectrum is unaltered. Above a certain $E_{\text{th}}$, however, the spectrum suddenly changes and two modes can be distinguished: m$_1$ (slow) and m$_2$ (fast). m$_1$ is the previously described soft mode (m$^*_{\text{NF,L}}$), which can be understood as the collective rotation of molecules around their short axis. The amplitude of this process should vary as $\Delta \varepsilon \propto (1+2S)$ \cite{kremer_broadband_2003, molecular_modes}, and the fitted amplitudes resemble the curves of Ref. \cite{nf_critical}. Thus, we believe that we are inducing the N$_{\text{F}}$ phase. The origin of m$_{2}$, on the contrary, is not so clear. Judging from its amplitude and relaxation frequency, it is likely to be of the same origin as m$_{\text{Iso}}$ (molecular or collective with short correlation length), and appears to be affected by the anisotropic environment, since its amplitude decreases quite rapidly with increasing bias field. Moreover, it might be the same mode as m$_{\text{NF,H}}$, which gets faster and of smaller amplitude as the phase becomes more ordered. In order to confirm the induction of the N$_{\text{F}}$ phase, we also performed SHG measurements under various DC electric fields. SHG is not allowed in media with inversion symmetry, so it is a powerful technique to probe polar structures. These experiments were done on a $9.5$ $\mu$m-thick $1$ mm-gap IPS cell with the polarization of the incident light parallel to the applied electric field in order to probe the longitudinal hyperpolarizability of UUQU-4-N. The results are shown in Fig. \ref{fig:shg}. It can be readily observed that, above a certain electric field, a strong SHG signal is detected whose functional shape greatly resembles that described in Ref. \cite{nf_critical}. It should be noted that the value of $E_{\text{th}}$ in this case appears to be slightly lower than in the dielectric measurements, but we attribute this to differences in the cell and electrode geometry.

\begin{figure}[h!]
\begin{center}
\includegraphics[width=0.65\textwidth]{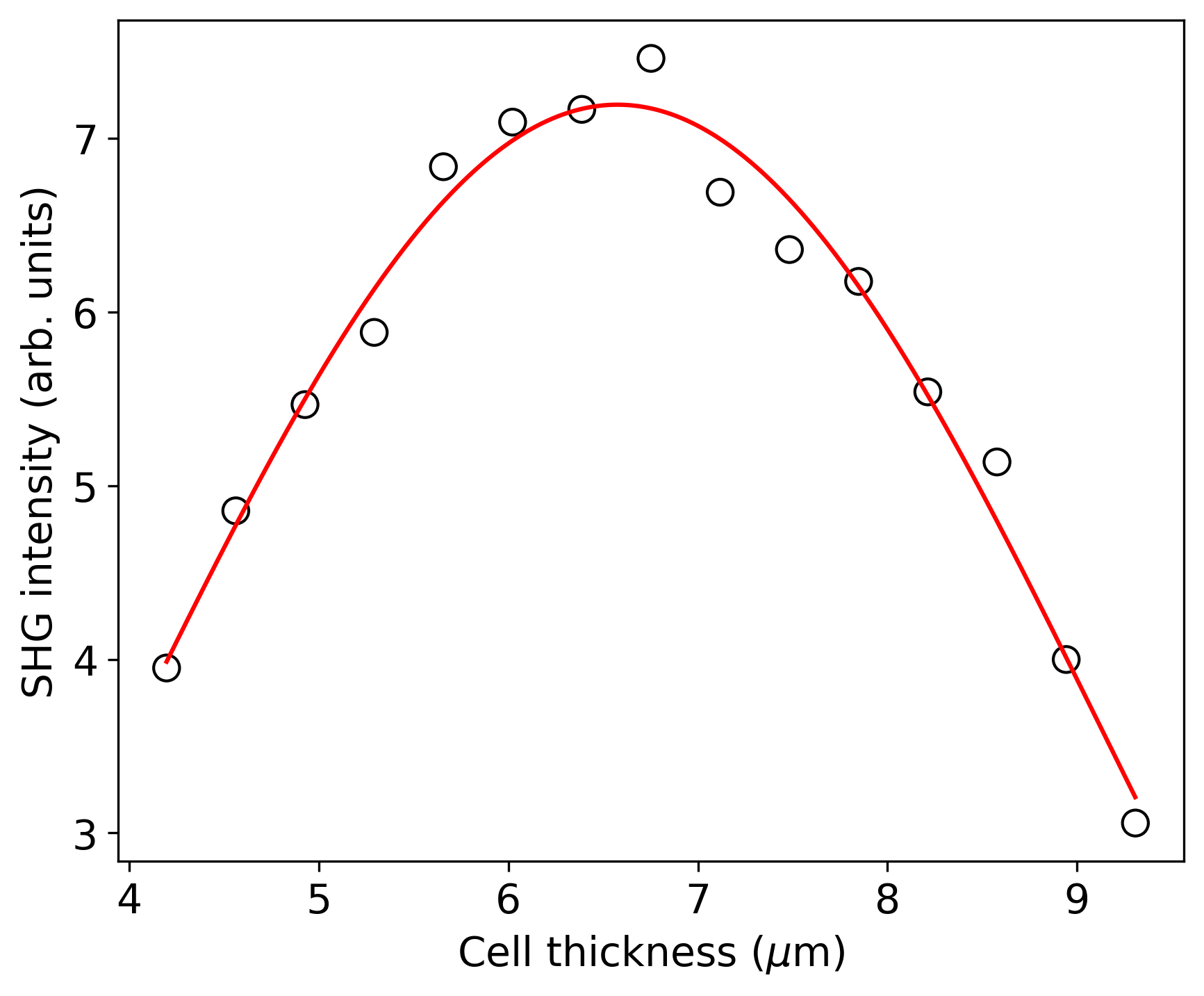}
\caption{\label{fig:shg_fit} Measured SHG intensity as a function of cell thickness at $18^{\circ}$C in the N$_{\text{F}}$ phase. The red line is a fit to Eq. \ref{shg}.}
\end{center}
\end{figure}

We have also characterized the SHG efficiency of UUQU-4-N in the N$_{\text{F}}$ phase, which allows for a direct connection with relevant molecular/structural parameters. Materials exhibiting the N$_{\text{F}}$ phase have been shown to present large nonlinear optical coefficients \cite{folcia_ferroelectric_2022}, not only because of the presence of strong donor-acceptor groups along their molecular long axis, but also due to the high degree of nematic order. The symmetry of the N$_{\text{F}}$ phase is $C_{\infty v}$, so that the second order dielectric susceptibility tensor (neglecting dispersion effects) can be expressed, under the supposition that Kleinman conditions hold, as

\begin{equation}
d =
    \begin{pmatrix}
    0 & 0 & 0 & 0 & d_{31} & 0\\
    0 & 0 & 0 & d_{31} & 0 & 0\\
    d_{31} & d_{31} & d_{33} & 0 & 0 & 0
    \end{pmatrix},\label{tensor}
\end{equation}
where the $Z$ axis has been taken to lie along the polar direction. Both $d_{33}$ and $d_{31}$ components can be determined by choosing the polarization of the incident light (extraordinary and ordinary, respectively). Due to the molecular structure of UUQU-4-N we have that $d_{33}\gg d_{31}$. To estimate the values we used a $0.5$ mm-thick $y$-cut quartz plate as reference. Dividing the SHG intensity of UUQU-4-N by that of quartz at the maximum of a Maker fringe, the data can be fitted to the following formula:

\begin{equation}
    \frac{I^{2\omega}_{\text{S}}}{I^{2\omega}_{\text{Q}}}=\frac{d^2_{\text{S}}(t^{2\omega}_{\text{S}})^2 (t^{\omega}_{\text{S}})^4 \Delta k_{\text{Q}}^2}{d^2_{\text{Q}}(t^{2\omega}_{\text{Q}})^2 (t^{\omega}_{\text{Q}})^4 \Delta k_{\text{S}}^2}\sin^2{\left(\Delta k_{\text{S}}\ell_{\text{S}}/2\right)},\label{shg}
\end{equation}
where $\Delta k = 4\pi \Delta n_{\text{d}}/\lambda$, $\Delta n_{\text{d}}=n^{2\omega}-n^{\omega}$ being the dispersion of the material, $\ell$ is the material thickness, and $t^{2\omega}$ and $t^{\omega}$ are the appropiate transmission Fresnel coefficients (the subscripts S and Q refer to the sample and quartz, respectively). The quartz data are $d_{\text{Q}}=0.4$ pm/V and $\Delta n_{\text{d}}=0.013$. Therefore, constructing a wedge cell with an appropriate thickness range and fitting the measured data to Eq. (\ref{shg}), $d_{33}$, $d_{31}$ and $\Delta n_{\text{d}}$ can in principle be determined. Selecting the polarization of the fundamental wave and outgoing SH wave along the polar axis, we obtained the results shown in Fig. \ref{fig:shg_fit}. Fitting the experimental data yields $d_{33}=4.3\pm 0.3$ pm/V and $\Delta n_{\text{d}}=0.05 \pm 0.01$. This value of $d_{33}$, although somewhat smaller than that of RM734 ($5.6$ pm/V) \cite{folcia_ferroelectric_2022}, is over an order of magnitude larger than that of DIO ($0.24$ pm/V) \cite{satoshi_shg}. $d_{31}$ could not be determined because, as argued before, it is much smaller than $d_{33}$ and we did not observe appreciable SHG signal in the appropiate polarizer configuration. 

In order to relate this quantity to molecular/structural parameters, the following expression can be used:

\begin{equation}
    d_{IJK}=N f^3 \langle \beta_{IJK} \rangle,
\end{equation}
where $N$ is the number of molecules per unit volume, $f=(n^2+2)/3$ is the Lorentz factor and $\langle \beta_{IJK} \rangle$ are the thermally averaged hyperpolarizability components in the laboratory frame ($Z$ is along the macroscopic polar axis). The expressions for the relevant components in terms of the molecular hyperpolarizabilities can be found in the SI (Eqs. (S1), (S2) and (S3)). One then obtains a final tensor, which has the shape given by Eq. (\ref{tensor}) in the two-index Voigt notation. It should be mentioned that the calculation of the hyperpolarizability tensor on the whole molecule usually yields inaccurate results because it is very sensitive to small changes in the molecular conformation. Therefore, we have assumed that the molecular hyperpolarizability comes essentially from the 4-(Difluoromethoxy)-2,6-difluorobenzonitrile segment. The results can be found in Table S3. Finally, we assume that the nematic potential at temperature $T$ can be described by this simple expression:

\begin{equation}
    U(\theta)=a(T)\sin^2{\theta}+b(T)\sin^2{(\theta/2)},
\end{equation}
where $a(T)$ is the strength of the nematic potential, while $b(T)$ accounts for the polar nature of the phase \cite{folcia_ferroelectric_2022}. The distribution function is then $F(\theta)=A\exp{[-U(\theta)/k_{\text{B}}T]}$, where $A$ is a normalization constant and $k_{\text{B}}$ is the Boltzmann constant. In order to determine $a(T)/k_{\text{B}}T$ and $b(T)/k_{\text{B}}T$ at $18^{\circ}$C, the following information will be considered: from the spontaneous polarization measured in Ref. \cite{manabe_eremin} ($\sim 5$ $\mu$C/cm$^2$) and taking $\rho=1.35$ g/cm$^3$, $M=0.4854$ kg/mol and $\mu=11.4$ D, $\langle P_1(\cos{\theta})\rangle \approx 0.8$; from the measured birefringence of UUQU-4-N in the N$_{\text{F}}$ phase \cite{manabe_eremin}, which is similar to that of RM734 \cite{mandle_molecular_2021}, $\langle P_{2}(\cos{\theta})\rangle \approx 0.7$ can be estimated. With these conditions, the numerically deduced values are $a(T)/k_{\text{B}}T=4.8$ and $b(T)/k_{\text{B}}T=3.5$. A plot of the nematic potential and distribution function can be found in Fig. S4. Thus, taking the refractive index $n=1.6$ and performing the corresponding thermal averages, we have $d_{33}=3.4$ pm/V and $d_{31}=0.1$ pm/V. With regards to $d_{33}$, the agreement between the experimental and theoretical values is rather remarkable but is only to be taken qualitatively. In the case of $d_{31}$, on the other hand, although no experimental value could be obtained, given the detection limit, it is reasonable to expect a value (at least) an order of magnitude smaller than $d_{33}$.

Having investigated the dynamical processes of UUQU-4-N, we shall address the emergence of ferroelectricity from a theoretical point of view. As explained in the Introduction, there are various interactions that could give rise to the N$_{\text{F}}$ phase. Perhaps the most obvious feature of ferroelectric nematogens is their strong dipole moment of the order of $10$ D. As already noted by Born in 1916 \cite{lavrentovich_review,born}, the dipole moment of the constituent molecules should be large enough to withstand the thermal fluctuations, namely $\mu^2/\varepsilon_0\varepsilon V>k_{\text{B}}T$, where $V$ is the molecular volume. This was supported by many theoretical works before the N$_{\text{F}}$ phase was discovered \cite{palffy-muhoray_ferroelectric_1988, wei_orientational_1992, lee_ferroelectric_1994}. As odd as it may seem, it was also backed by some experiments. For instance, a polar (and highly viscous) N phase was found in some aromatic copolyesters with a high degree of polymerization $X_n$ \cite{Watanabe_1996}. Since the compounds assumed an extended conformation, their dipole moment was proportional to $X_n$. As the molar volume is also proportional to $X_n$, the aforementioned argument would predict that only compounds with a dipole strength $\mu^* \propto \sqrt{\mu^2/V} \propto \sqrt{X_n}$ exceeding a critical value would form the polar phase. Indeed, the authors observed a SHG signal only for compounds exceeding a certain inherent viscosity, which is related to $X_n$ and, thus, $\mu^*$. This constituted an indirect evidence that dipole-dipole interactions gave rise to the polar N phase. However, now that the existence of the N$_{\text{F}}$ phase has been realized in low-molar-mass LCs, this simplified picture has turned out to be incomplete. Even though a critical dipole moment ($\sim 9$ D \cite{Nishikawa_giant}) does seem to exist, a number of molecules whose dipole moment exceeds this value do not exhibit the N$_{\text{F}}$ phase \cite{Nishikawa_giant, li_general_2022}. A especially intriguing example is RM734 and its analog RM734-CN: when the terminal nitro group is replaced by a cyano group, the polar phase disappears \cite{mandle_molecular_2021}. On the other hand, the concepts of excluded volume and packing entropy have been widely exploited and are thought to be responsible for the structure of many mesophases \cite{goodby_volume, zannoni_book}. This might also be the case for N$_{\text{F}}$ materials, since the presence of a steric dipole in some ferroelectric nematogens like RM734 could help to stabilize the polar structure. Although a DFT-based approach developed a few years ago seemed promising \cite{dft_ns}, Monte Carlo simulations have been unable to replicate such results \cite{dft_vs_mc_2023, tapered_2023}.

\begin{figure}[h!]
\begin{center}
\includegraphics[width=0.7\textwidth]{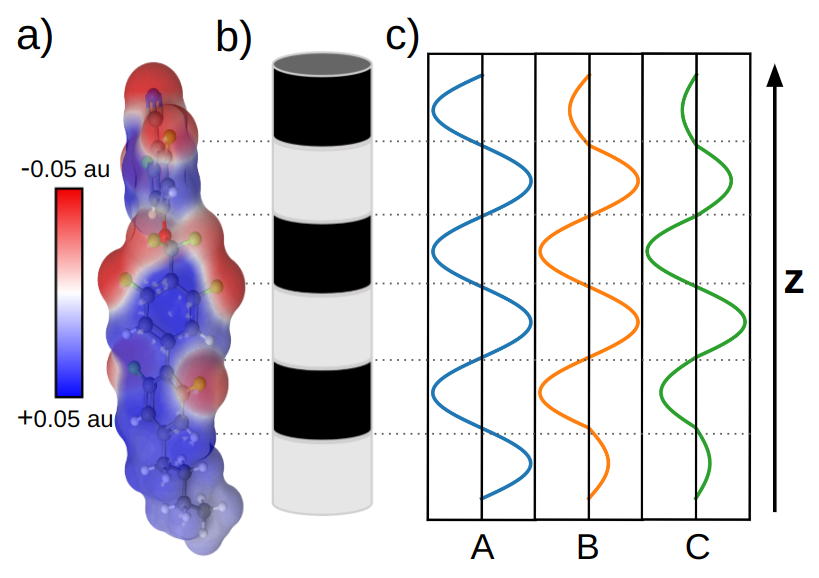}
\caption{\label{fig:model_1} Modeling of the UUQU-4-N molecule: a) molecular electrostatic potential (MEP) mapped onto the Van der Waals surface with the DFT-optimized geometry. b) The molecule is modeled as a cylindrical rod bearing three longitudinal charge density waves of wavelength $\lambda = l/3$. The colors indicate alternating positive and negative charge. c) The charge density is assumed to have a sinusoidal variation along its long axis. The graph shows the cases studied with different amplitudes of neighboring half-waves: A with $s_i=1$ ($i=1$--$6$); B with $s_1=s_6=0.4$ and $s_2=...=s_5=1$; C with $s_1=s_6=0.4$, $s_3=s_4=1.4$ and $s_2=s_5=1$.}
\end{center}
\end{figure}

With the aim of finding the mechanism for the formation of the N$_{\text{F}}$ phase, Madhusudana recently developed a simple molecular model purely on the basis of short-range electrostatic interactions \cite{madhusudana}. The ferroelectric nematogens are modeled as cylindrical rods bearing four longitudinal surface charge density waves of wavelength $\lambda=l/4$, $l$ being the molecular length. In fact, this seems to be a common feature in most ferroelectric nematic materials. Additionally, the charge density is assumed to vary sinusoidally along the length of the molecule, but the amplitudes of neighboring half-waves (denoted by $s_i$) need not be equal. The interaction between two molecules is then modeled as the sum of electrostatic interactions between all half-waves according to Eqs. (2), (3) and (4) in Ref. \cite{madhusudana}. According to the calculations, the most favorable case for the formation of the N$_{\text{F}}$ phase is when the amplitudes of the terminal half-waves are reduced and the inner ones are enhanced. Despite the simplicity of this model, which only focuses on electrostatics, it not only successfully predicts the stability of the N$_{\text{F}}$ phase but explains why certain molecules do not exhibit such phase based on their charge distribution. The author argues, for instance, that the aforementioned RM734-CN does not exhibit the polar phase because the linear cyano group has a higher charge density than the nitro group. Other authors have also found qualitative agreement with Madhusudana's model \cite{lateral_chain_effect, gun_shape_nf, shakespeare_nf, sulfur_nf}. In one of these works, for example, the authors enhance the stability of the N$_{\text{F}}$ phase in a certain molecule by adding a fluorine atom \textit{ortho} to the terminal nitro group, which helps to spread the charge at the end \cite{gun_shape_nf}. Here we explore the applicability of Madhusudana's model in the case of UUQU-4-N. To this end, we have first computed its molecular electrostatic potential (MEP), as shown in Fig. \ref{fig:model_1}a. We can clearly see that it exhibits regions of alternating positive (blue) and negative (red) charge. However, unlike in most ferroelectric nematic materials, it only shows six clearly differentiated regions. Thus, we model the molecule as a cylindrical rod with three longitudinal surface charge density waves of wavelength $\lambda=l/3$  (see Figs. \ref{fig:model_1}b and \ref{fig:model_1}c). We used $l=22$ \AA{}, while the radius of the rod was set to $r=2$ \AA{}. A first approach involves calculating the interaction energy between two rods in the parallel and antiparallel configurations as a function of the relative shift $\zeta$ between the two rods (see Fig. S5). We have considered three different possibilities for the amplitudes of the charge density waves: cases A (equal amplitudes for all waves), B (reduced amplitudes at the ends) and C (reduced amplitudes at the ends and enhanced in the middle) depicted in Fig. \ref{fig:model_1}.

\begin{figure}[h!]
\begin{center}
\includegraphics[width=0.75\textwidth]{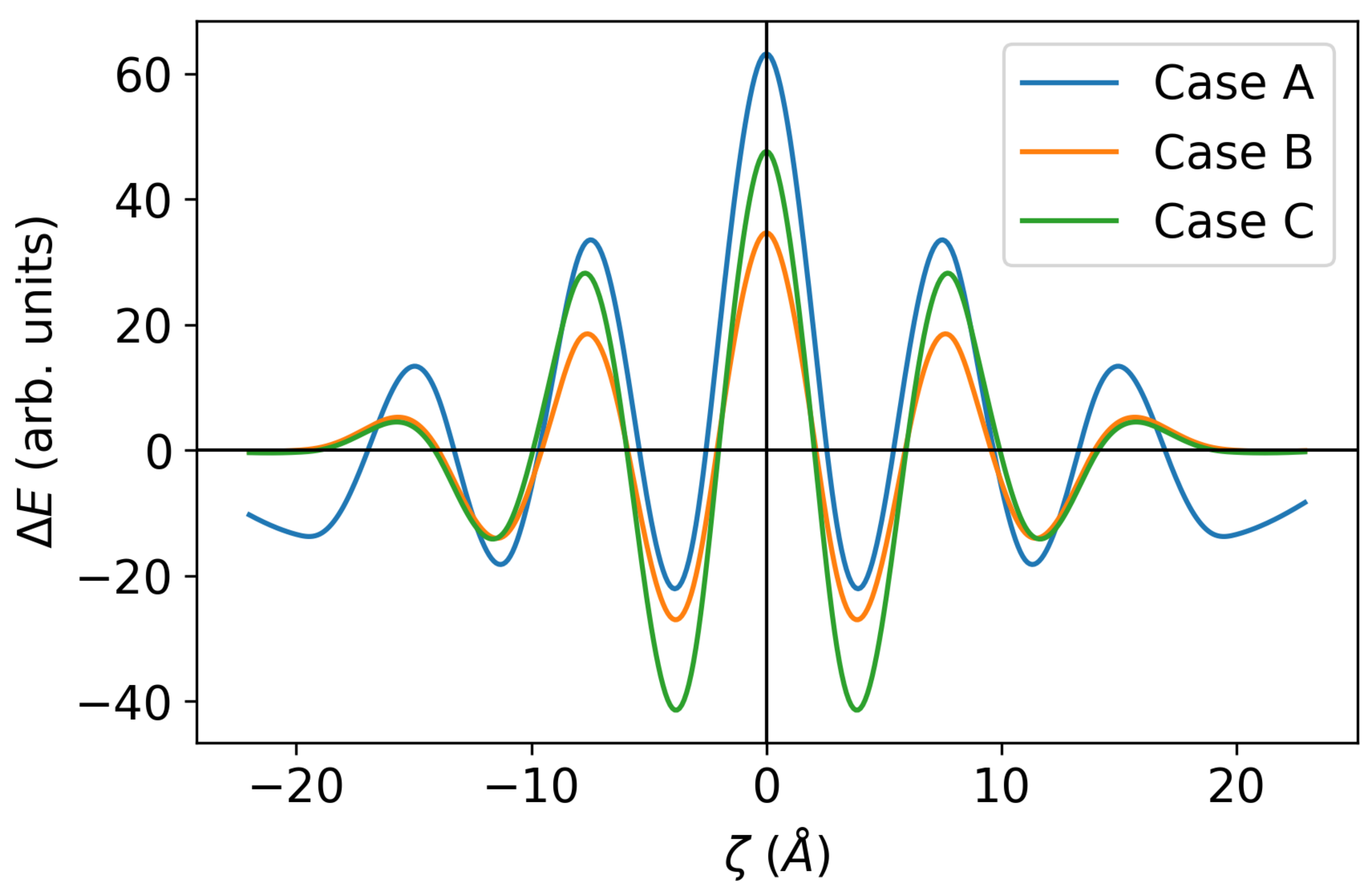}
\caption{\label{fig:model_2} Difference in energy ($\Delta E$) between parallel and antiparallel rods as a function of the shift $\zeta$ for $R_0=4.5$ \AA{} obtained with the three different charge distributions have been considered (see Fig. \ref{fig:model_1})}
\end{center}
\end{figure}

Setting the inter-rod separation at $R_0=4.5$ \AA{}, as inferred from our XRD data, we analyze how the energy difference between the parallel and antiparallel configurations $\Delta E=E_{\text{P}}-E_{\text{AP}}$ varies with the relative shift $\zeta$. The results is shown in Fig. \ref{fig:model_2}. As expected, at $\zeta=0$ the antiparallel configuration is favored. However, when the shift is such that the half-waves of opposite charges align, the parallel configuration is favored. A total of 6 minima appear in each curve located approximately at multiples of $l/6$ due to the 6 regions of alternating charge. In case A, there is a big difference in energy between the central maximum and the first minimum, so we conclude that the N$_{\text{F}}$ phase cannot form. If we reduce the amplitudes of the outer half-waves as in case B, and simultaneously enhance those of the inner ones as in case C, the situation is more favorable. This latter case may well represent the UUQU-4-N molecule better since its relatively short alkyl chain and spatially distributed polar groups reduce their respective charge densities at the ends, while the strongly electronegative nature of the oxygen and fluorine atoms at the center may enhance the inner half-waves.

\begin{figure}
\begin{center}
\includegraphics[width=0.65\textwidth]{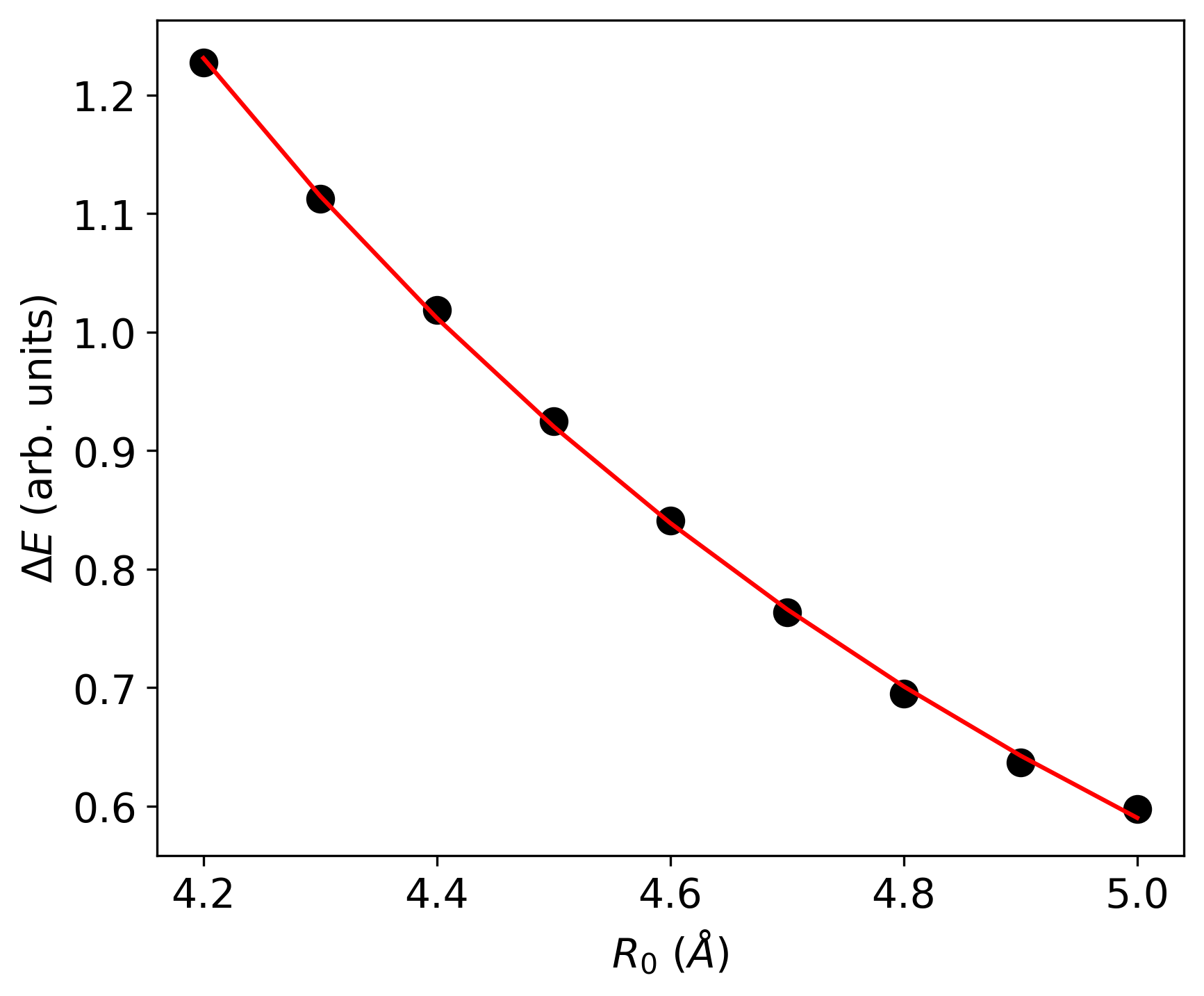}
\caption{\label{fig:model_3} Difference in energy ($\Delta E$) between the ferroelectric and antiferroelectric structure (the red line is a guide for the eye).}
\end{center}
\end{figure}

It is evident that calculations involving two rods, although interesting as a first approach, can hardly represent the real N$_{\text{F}}$ phase. Following Madushudana's work, we run the calculations for case C using a small two-layer cluster with a total of 38 rods giving rise to a total of 703 inter-rod interactions (see details and assumptions in the SI). We minimized the total energy for the ferroelectric and antiferroelectric structures with the results of the calculations shown in Fig. \ref{fig:model_3}. The graph represents the energy difference between structures, $\Delta E = E_{\text{F}}-E_{\text{AF}}$, as a function of $R_0$. We see that the N$_{\text{F}}$ phase is not stable for any value of $R_0$. Thus, the model predicts a conventional Iso--N phase transition. On the contrary, in the case of rods bearing $4$ charge density waves, there existed a certain range of $R_0$ (compatible with the distance associated with the diffuse halo from XRD measurements) for which the ferroelectric phase was favored \cite{madhusudana}.

Rationalizing these findings in terms of the UUQU-4-N molecule, there are a couple of reasons that could explain the aforementioned results. One of them is that, even though electrostatic interactions are behind the origin of the N$_{\text{F}}$ phase in this system, finer details of the molecular structure and charge distribution are not taken into account in such a coarse-grained model. This assertion derived from our results can be compared with a set of atomistic MD simulations that were performed for this material by Richard J. Mandle (see Ref. \cite{manabe_md}). It is important to point out that, in such simulations, given a molecular structure, the N$_{\text{F}}$ phase does not spontaneously arise and it must be induced by hand, hence, the interactions that govern the ferroelectric transition can only be indirectly inferred. First of all, Madhusudana's assumption that two pseudolayers suffice to accurately describe the ferroelectric phase seems justified, since Mandle notes that, at separations corresponding to the molecular length along the $z$ axis, the dipole-dipole interaction strength is small and not much larger than $k_{\text{B}}T$. Secondly, Mandle explains the favorable effect of short alkyl chains to form the N$_{\text{F}}$ phase by a greater molecular packing, unlike Madhusudana who interprets it as an electrostatic effect. Lastly, the parallel side-by-side molecular arrangement appears to be driven by the electrostatic interaction of fluorine atoms with hydrogens. This is indeed the evidence in favor of a more detailed electrostatic interaction scenario. On the other hand, excluded volume interactions could definitely help the formation of the N$_{\text{F}}$ phase. Nevertheless, it is difficult to ground this on simple physical terms. If the UUQU-4-N molecule had a more pronounced wedged shape, for instance, it would be easy to argue that it would help the formation of a polar N$_{\text{S}}$ phase. In any case, it should be noted that UUQU-4-N is a rather peculiar system in which the polar order only appears on supercooling, at a temperature much lower than the Cr--Iso transition on heating, which poses serious doubts on its stability. Thus, the emergence of polar order in this system probably relies on a subtle balance between electrostatics and molecular shape.

\section{Conclusions}
In this work, we have attempted to answer how spontaneous polar order emerges in a compound directly below the isotropic liquid state. Broadband dielectric spectra reveal that polar correlations start to occur already in the Iso phase, whose magnitude presumably grows while approaching the Iso--N$_{\text{F}}$ phase transition. In fact, the ferroelectric phase can be induced close to the transition temperature by applying a sufficiently strong electric field, as confirmed by second harmonic generation measurements. These measurements have also allowed us to estimate the main coefficient of the second order dielectric susceptibility tensor in the N$_{\text{F}}$ phase, obtaining a value of $d_{33}\approx 4.3$ pm/V. We have also ruled out the possibility that the N$_{\text{F}}$ phase in general exhibits a colossal dielectric constant by a careful analysis of dielectric data, estimating a value for the parallel component of this quantity of $\varepsilon_{\parallel} \approx 150$. Lastly, calculations based on a recently proposed theoretical model focusing on intermolecular electrostatic interactions suggest that the formation of the ferroelectric phase in this system cannot be explained solely on the basis of surface charge density waves as proposed by Madhusudana.

\section*{Acknowledgements}
The ferroelectric nematic liquid crystal material used in this work was supplied by Merck Electronics KGaA. A.E. and J.M.-P. acknowledge funding from the Basque Goverment Project IT1458-22, as well as the technical and human support provided by IZO-SGI SGIker of UPV/EHU and European funding (ERDF and ESF). A.E. thanks the Department of Education of the Basque Government for a predoctoral fellowship (grant number PRE\_2023\_2\_0113). N.S and A.M acknowledge financial support from the Slovenian Research and Innovation Agency (ARIS) (grant numbers P1-0192 and J1-50004). 

\appendix

\bibliographystyle{elsarticle-num} 
\bibliography{REFS}

\begin{thebibliography}{10}
\expandafter\ifx\csname url\endcsname\relax
  \def\url#1{\texttt{#1}}\fi
\expandafter\ifx\csname urlprefix\endcsname\relax\def\urlprefix{URL }\fi
\expandafter\ifx\csname href\endcsname\relax
  \def\href#1#2{#2} \def\path#1{#1}\fi

\bibitem{mandle_nematic_2017}
R.~J. Mandle, S.~J. Cowling, J.~W. Goodby, A nematic to nematic transformation exhibited by a rod-like liquid crystal, Physical Chemistry Chemical Physics 19~(18) (2017) 11429--11435.
\newblock \href {https://doi.org/10.1039/C7CP00456G} {\path{doi:10.1039/C7CP00456G}}.

\bibitem{nishikawa_fluid_2017}
H.~Nishikawa, K.~Shiroshita, H.~Higuchi, Y.~Okumura, Y.~Haseba, S.-i. Yamamoto, K.~Sago, H.~Kikuchi, A {Fluid} {Liquid}-{Crystal} {Material} with {Highly} {Polar} {Order}, Advanced Materials 29~(43) (2017) 1702354.
\newblock \href {https://doi.org/10.1002/adma.201702354} {\path{doi:10.1002/adma.201702354}}.

\bibitem{chen_first-principles_2020}
X.~Chen, E.~Korblova, D.~Dong, X.~Wei, R.~Shao, L.~Radzihovsky, M.~A. Glaser, J.~E. Maclennan, D.~Bedrov, D.~M. Walba, N.~A. Clark, First-principles experimental demonstration of ferroelectricity in a thermotropic nematic liquid crystal: {Polar} domains and striking electro-optics, Proceedings of the National Academy of Sciences 117~(25) (2020) 14021--14031.
\newblock \href {https://doi.org/10.1073/pnas.2002290117} {\path{doi:10.1073/pnas.2002290117}}.

\bibitem{lavrentovich_review}
O.~D. Lavrentovich, Ferroelectric nematic liquid crystal, a century in waiting, Proceedings of the National Academy of Sciences 117~(26) (2020) 14629–14631.
\newblock \href {https://doi.org/10.1073/pnas.2008947117} {\path{doi:10.1073/pnas.2008947117}}.

\bibitem{nf_review}
N.~Sebasti\'an, M.~\ifmmode \check{C}\else \v{C}\fi{}opi\ifmmode~\check{c}\else \v{c}\fi{}, A.~Mertelj, Ferroelectric nematic liquid-crystalline phases, Phys. Rev. E 106 (2022) 021001.
\newblock \href {https://doi.org/10.1103/PhysRevE.106.021001} {\path{doi:10.1103/PhysRevE.106.021001}}.

\bibitem{brown_multiple_2021}
S.~Brown, E.~Cruickshank, J.~M.~D. Storey, C.~T. Imrie, D.~Pociecha, M.~Majewska, A.~Makal, E.~Gorecka, Multiple {Polar} and {Non}‐polar {Nematic} {Phases}, ChemPhysChem 22~(24) (2021) 2506--2510.
\newblock \href {https://doi.org/10.1002/cphc.202100644} {\path{doi:10.1002/cphc.202100644}}.

\bibitem{folcia_ferroelectric_2022}
C.~L. Folcia, J.~Ortega, R.~Vidal, T.~Sierra, J.~Etxebarria, The ferroelectric nematic phase: an optimum liquid crystal candidate for nonlinear optics, Liquid Crystals 49~(6) (2022) 899--906.
\newblock \href {https://doi.org/10.1080/02678292.2022.2056927} {\path{doi:10.1080/02678292.2022.2056927}}.

\bibitem{li_how_2021}
J.~Li, R.~Xia, H.~Xu, J.~Yang, X.~Zhang, J.~Kougo, H.~Lei, S.~Dai, H.~Huang, G.~Zhang, F.~Cen, Y.~Jiang, S.~Aya, M.~Huang, How {Far} {Can} {We} {Push} the {Rigid} {Oligomers}/{Polymers} toward {Ferroelectric} {Nematic} {Liquid} {Crystals}?, Journal of the American Chemical Society 143~(42) (2021) 17857--17861.
\newblock \href {https://doi.org/10.1021/jacs.1c09594} {\path{doi:10.1021/jacs.1c09594}}.

\bibitem{Nishikawa_giant}
J.~Li, H.~Nishikawa, J.~Kougo, J.~Zhou, S.~Dai, W.~Tang, X.~Zhao, Y.~Hisai, M.~Huang, S.~Aya, Development of ferroelectric nematic fluids with giant-$\varepsilon$ dielectricity and nonlinear optical properties, Science Advances 7~(17) (2021) eabf5047.
\newblock \href {https://doi.org/10.1126/sciadv.abf5047} {\path{doi:10.1126/sciadv.abf5047}}.

\bibitem{ferroelastic}
N.~Sebasti\'an, L.~Cmok, R.~J. Mandle, M.~R. de~la Fuente, I.~Dreven\ifmmode \check{s}\else~\v{s}\fi{}ek Olenik, M.~\ifmmode \check{C}\else \v{C}\fi{}opi\ifmmode~\check{c}\else \v{c}\fi{}, A.~Mertelj, Ferroelectric-ferroelastic phase transition in a nematic liquid crystal, Physical Review Letters 124 (2020) 037801.
\newblock \href {https://doi.org/10.1103/PhysRevLett.124.037801} {\path{doi:10.1103/PhysRevLett.124.037801}}.

\bibitem{chen_smectic_2023}
X.~Chen, V.~Martinez, E.~Korblova, G.~Freychet, M.~Zhernenkov, M.~A. Glaser, C.~Wang, C.~Zhu, L.~Radzihovsky, J.~E. Maclennan, D.~M. Walba, N.~A. Clark, {The smectic {Z}$_{\textrm{{A}}}$ phase: {Antiferroelectric} smectic order as a prelude to the ferroelectric nematic}, Proceedings of the National Academy of Sciences 120~(8) (2023) e2217150120.
\newblock \href {https://doi.org/10.1073/pnas.2217150120} {\path{doi:10.1073/pnas.2217150120}}.

\bibitem{chen_smectic_a}
X.~Chen, V.~Martinez, P.~Nacke, E.~Korblova, A.~Manabe, M.~Klasen-Memmer, G.~Freychet, M.~Zhernenkov, M.~A. Glaser, L.~Radzihovsky, J.~E. Maclennan, D.~M. Walba, M.~Bremer, F.~Giesselmann, N.~A. Clark, Observation of a uniaxial ferroelectric smectic a phase, Proceedings of the National Academy of Sciences 119~(47) (Nov. 2022).
\newblock \href {https://doi.org/10.1073/pnas.2210062119} {\path{doi:10.1073/pnas.2210062119}}.

\bibitem{ferroelectric_applications}
L.~W. Martin, A.~M. Rappe, Thin-film ferroelectric materials and their applications, Nat. Rev. Mater. 2~(2) (Nov. 2016).

\bibitem{palffy-muhoray_ferroelectric_1988}
P.~Palffy-Muhoray, M.~A. Lee, R.~G. Petschek, Ferroelectric {Nematic} {Liquid} {Crystals}: {Realizability} and {Molecular} {Constraints}, Physical Review Letters 60~(22) (1988) 2303--2306.
\newblock \href {https://doi.org/10.1103/PhysRevLett.60.2303} {\path{doi:10.1103/PhysRevLett.60.2303}}.

\bibitem{biscarini_head-tail_nodate}
C.~C. F.~Biscarini, C.~Zannoni, P.~Pasini, Head-tail asymmetry and ferroelectricity in uniaxial liquid crystals, Molecular Physics 73~(2) (1991) 439--461.
\newblock \href {https://doi.org/10.1080/00268979100101301} {\path{doi:10.1080/00268979100101301}}.

\bibitem{wei_orientational_1992}
D.~Wei, G.~N. Patey, Orientational order in simple dipolar liquids: {Computer} simulation of a ferroelectric nematic phase, Physical Review Letters 68~(13) (1992) 2043--2045.
\newblock \href {https://doi.org/10.1103/PhysRevLett.68.2043} {\path{doi:10.1103/PhysRevLett.68.2043}}.

\bibitem{lee_ferroelectric_1994}
J.~Lee, S.-D. Lee, Ferroelectric {Liquid} {Crystalline} {Ordering} of {Rigid} {Rods} with {Dipolar} {Interactions}, Molecular Crystals and Liquid Crystals Science and Technology. Section A. Molecular Crystals and Liquid Crystals 254~(1) (1994) 395--403.
\newblock \href {https://doi.org/10.1080/10587259408036088} {\path{doi:10.1080/10587259408036088}}.

\bibitem{berardi_ferroelectric_2001}
R.~Berardi, M.~Ricci, C.~Zannoni, Ferroelectric {Nematic} and {Smectic} {Liquid} {Crystals} from {Tapered} {Molecules}, ChemPhysChem 2~(7) (2001) 443--447.
\newblock \href {https://doi.org/10.1002/1439-7641(20010716)2:7<443::AID-CPHC443>3.0.CO;2-J} {\path{doi:10.1002/1439-7641(20010716)2:7<443::AID-CPHC443>3.0.CO;2-J}}.

\bibitem{yadav_polar_2022}
N.~Yadav, Y.~P. Panarin, J.~K. Vij, W.~Jiang, G.~H. Mehl, Two mechanisms for the formation of the ferronematic phase studied by dielectric spectroscopy, Journal of Molecular Liquids 378 (2023) 121570.
\newblock \href {https://doi.org/10.1016/j.molliq.2023.121570} {\path{doi:10.1016/j.molliq.2023.121570}}.

\bibitem{erkoreka_rm734}
A.~Erkoreka, J.~Martinez-Perdiguero, R.~J. Mandle, A.~Mertelj, N.~Sebastián, Dielectric spectroscopy of a ferroelectric nematic liquid crystal and the effect of the sample thickness, Journal of Molecular Liquids 387 (2023) 122566.
\newblock \href {https://doi.org/10.1016/j.molliq.2023.122566} {\path{doi:10.1016/j.molliq.2023.122566}}.

\bibitem{erkoreka_dio}
A.~Erkoreka, A.~Mertelj, M.~Huang, S.~Aya, N.~Sebastián, J.~Martinez-Perdiguero, {Collective and non-collective molecular dynamics in a ferroelectric nematic liquid crystal studied by broadband dielectric spectroscopy}, The Journal of Chemical Physics 159~(18) (2023) 184502.
\newblock \href {https://doi.org/10.1063/5.0173813} {\path{doi:10.1063/5.0173813}}.

\bibitem{dft_ns}
P.~De~Gregorio, E.~Frezza, C.~Greco, A.~Ferrarini, Density functional theory of nematic elasticity: softening from the polar order, Soft Matter 12 (2016) 5188--5198.
\newblock \href {https://doi.org/10.1039/C6SM00624H} {\path{doi:10.1039/C6SM00624H}}.

\bibitem{mandle_molecular_2021}
R.~J. Mandle, N.~Sebastián, J.~Martinez-Perdiguero, A.~Mertelj, On the molecular origins of the ferroelectric splay nematic phase, Nature Communications 12~(1) (2021) 4962.
\newblock \href {https://doi.org/10.1038/s41467-021-25231-0} {\path{doi:10.1038/s41467-021-25231-0}}.

\bibitem{dft_vs_mc_2023}
P.~Kubala, M.~Cieśla, Splay and polar order in a system of hard pear-like molecules: confrontation of monte carlo numerical simulations with density functional theory calculations, Soft Matter 19 (2023) 7836--7845.
\newblock \href {https://doi.org/10.1039/D3SM01021J} {\path{doi:10.1039/D3SM01021J}}.

\bibitem{tapered_2023}
P.~Kubala, M.~Cie\ifmmode~\acute{s}\else \'{s}\fi{}la, L.~Longa, Splay-induced order in systems of hard tapers, Phys. Rev. E 108 (2023) 054701.
\newblock \href {https://doi.org/10.1103/PhysRevE.108.054701} {\path{doi:10.1103/PhysRevE.108.054701}}.

\bibitem{manabe_md}
R.~J. Mandle, In silico interactome of a room-temperature ferroelectric nematic material, Crystals 13~(6) (2023).
\newblock \href {https://doi.org/10.3390/cryst13060857} {\path{doi:10.3390/cryst13060857}}.

\bibitem{madhusudana}
N.~V. Madhusudana, Simple molecular model for ferroelectric nematic liquid crystals exhibited by small rodlike mesogens, Phys. Rev. E 104 (2021) 014704.
\newblock \href {https://doi.org/10.1103/PhysRevE.104.014704} {\path{doi:10.1103/PhysRevE.104.014704}}.

\bibitem{manabe_ferroelectric_2021}
A.~Manabe, M.~Bremer, M.~Kraska, Ferroelectric nematic phase at and below room temperature, Liquid Crystals 48~(8) (2021) 1079--1086.
\newblock \href {https://doi.org/10.1080/02678292.2021.1921867} {\path{doi:10.1080/02678292.2021.1921867}}.

\bibitem{manabe_eremin}
E.~Zavvou, M.~Klasen-Memmer, A.~Manabe, M.~Bremer, A.~Eremin, Polarisation-driven magneto-optical and nonlinear-optical behaviour of a room-temperature ferroelectric nematic phase, Soft Matter 18 (2022) 8804--8812.
\newblock \href {https://doi.org/10.1039/D2SM01298G} {\path{doi:10.1039/D2SM01298G}}.

\bibitem{manabe_original}
A.~Manabe, M.~Bremer, M.~Kraska, Ferroelectric nematic phase at and below room temperature, Liquid Crystals 48~(8) (2021) 1079--1086.
\newblock \href {https://doi.org/10.1080/02678292.2021.1921867} {\path{doi:10.1080/02678292.2021.1921867}}.

\bibitem{vaupotic_dielectric_2023}
N.~Vaupotič, D.~Pociecha, P.~Rybak, J.~Matraszek, M.~Čepič, J.~M. Wolska, E.~Gorecka, Dielectric response of a ferroelectric nematic liquid crystalline phase in thin cells, Liquid Crystals (2023) 1--12\href {https://doi.org/10.1080/02678292.2023.2180099} {\path{doi:10.1080/02678292.2023.2180099}}.

\bibitem{clark_dielectric_2022}
N.~A. Clark, X.~Chen, J.~E. MacLennan, M.~A. Glaser, Dielectric spectroscopy of ferroelectric nematic liquid crystals: Measuring the capacitance of insulating interfacial layers, Phys. Rev. Res. 6 (2024) 013195.
\newblock \href {https://doi.org/10.1103/PhysRevResearch.6.013195} {\path{doi:10.1103/PhysRevResearch.6.013195}}.

\bibitem{orca}
F.~Neese, {Software update: The ORCA program system—Version 5.0}, WIREs Computational Molecular Science 12~(5) (2022) e1606.
\newblock \href {https://doi.org/https://doi.org/10.1002/wcms.1606} {\path{doi:https://doi.org/10.1002/wcms.1606}}.

\bibitem{g16}
M.~J. Frisch, G.~W. Trucks, H.~B. Schlegel, G.~E. Scuseria, M.~A. Robb, J.~R. Cheeseman, G.~Scalmani, V.~Barone, G.~A. Petersson, H.~Nakatsuji, X.~Li, M.~Caricato, A.~V. Marenich, J.~Bloino, B.~G. Janesko, R.~Gomperts, B.~Mennucci, H.~P. Hratchian, J.~V. Ortiz, A.~F. Izmaylov, J.~L. Sonnenberg, D.~Williams-Young, F.~Ding, F.~Lipparini, F.~Egidi, J.~Goings, B.~Peng, A.~Petrone, T.~Henderson, D.~Ranasinghe, V.~G. Zakrzewski, J.~Gao, N.~Rega, G.~Zheng, W.~Liang, M.~Hada, M.~Ehara, K.~Toyota, R.~Fukuda, J.~Hasegawa, M.~Ishida, T.~Nakajima, Y.~Honda, O.~Kitao, H.~Nakai, T.~Vreven, K.~Throssell, J.~A. Montgomery, {Jr.}, J.~E. Peralta, F.~Ogliaro, M.~J. Bearpark, J.~J. Heyd, E.~N. Brothers, K.~N. Kudin, V.~N. Staroverov, T.~A. Keith, R.~Kobayashi, J.~Normand, K.~Raghavachari, A.~P. Rendell, J.~C. Burant, S.~S. Iyengar, J.~Tomasi, M.~Cossi, J.~M. Millam, M.~Klene, C.~Adamo, R.~Cammi, J.~W. Ochterski, R.~L. Martin, K.~Morokuma, O.~Farkas, J.~B. Foresman, D.~J. Fox, Gaussian 16 {R}evision {C}.01, {Gaussian Inc.
  Wallingford CT} (2016).

\bibitem{jmol}
{Jmol: an open-source Java viewer for chemical structures in 3D}, \url{http://www.jmol.org/}.

\bibitem{kremer_broadband_2003}
F.~Kremer, A.~Schönhals (Eds.), Broadband {Dielectric} {Spectroscopy}, Springer Berlin Heidelberg, Berlin, Heidelberg, 2003.
\newblock \href {https://doi.org/10.1007/978-3-642-56120-7} {\path{doi:10.1007/978-3-642-56120-7}}.

\bibitem{7cb_dipole}
E.~M. K.~P.~Gueu, A.~Proutiere, {Dipole Moments of 4-n Alkyl-4'-Cyanobiphenyl Molecules (from OCB to 12CB) Measurement in Four Solvents and Theoretical Calculations}, Molecular Crystals and Liquid Crystals 132~(3-4) (1986) 303--323.
\newblock \href {https://doi.org/10.1080/00268948608079550} {\path{doi:10.1080/00268948608079550}}.

\bibitem{haase_relaxation_2003}
W.~Haase, S.~Wróbel (Eds.), Relaxation {Phenomena}, Springer Berlin Heidelberg, Berlin, Heidelberg, 2003.
\newblock \href {https://doi.org/10.1007/978-3-662-09747-2} {\path{doi:10.1007/978-3-662-09747-2}}.

\bibitem{Jackson}
J.~D. Jackson, Classical Electrodynamics, 3rd Edition, John Wiley \& Sons, Nashville, TN, 1998.

\bibitem{induced_nematic}
I.~Lelidis, G.~Durand, Electric-field-induced isotropic-nematic phase transition, Phys. Rev. E 48 (1993) 3822--3824.
\newblock \href {https://doi.org/10.1103/PhysRevE.48.3822} {\path{doi:10.1103/PhysRevE.48.3822}}.

\bibitem{nf_critical}
J.~Szydlowska, P.~Majewski, M.~\ifmmode \check{C}\else \v{C}\fi{}epi\ifmmode~\check{c}\else \v{c}\fi{}, N.~c.~v. Vaupoti\ifmmode~\check{c}\else \v{c}\fi{}, P.~Rybak, C.~T. Imrie, R.~Walker, E.~Cruickshank, J.~M.~D. Storey, P.~Damian, E.~Gorecka, Ferroelectric nematic-isotropic liquid critical end point, Phys. Rev. Lett. 130 (2023) 216802.
\newblock \href {https://doi.org/10.1103/PhysRevLett.130.216802} {\path{doi:10.1103/PhysRevLett.130.216802}}.

\bibitem{molecular_modes}
J.~Jadżyn, G.~Czechowski, R.~Douali, C.~Legrand, On the molecular interpretation of the dielectric relaxation of nematic liquid crystals, Liquid Crystals 26~(11) (1999) 1591–1597.
\newblock \href {https://doi.org/10.1080/026782999203571} {\path{doi:10.1080/026782999203571}}.

\bibitem{satoshi_shg}
R.~Xia, X.~Zhao, J.~Li, H.~Lei, Y.~Song, W.~Peng, X.~Zhang, S.~Aya, M.~Huang, Achieving enhanced second-harmonic generation in ferroelectric nematics by doping d–$\pi$–a chromophores, J. Mater. Chem. C 11 (2023) 10905--10910.
\newblock \href {https://doi.org/10.1039/D3TC01384G} {\path{doi:10.1039/D3TC01384G}}.

\bibitem{born}
M.~Born, {Über anisotrope Flüssigkeiten: Versuch einer Theorie der flüssigen Kristalle und des elektrischen Kerr-Effekts in Flüssigkeiten}, Sitzungsberichte der Preussischen Akademie der Wissenschaften 30 (1916) 614--650.

\bibitem{Watanabe_1996}
T.~Watanabe, S.~Miyata, T.~Furukawa, H.~Takezoe, T.~Nishi, M.~Sone, A.~Migita, J.~Watanabe, Nematic liquid crystals with polar ordering formed from simple aromatic polyester, Japanese Journal of Applied Physics 35~(4B) (1996) L505.
\newblock \href {https://doi.org/10.1143/JJAP.35.L505} {\path{doi:10.1143/JJAP.35.L505}}.

\bibitem{li_general_2022}
J.~Li, Z.~Wang, M.~Deng, Y.~Zhu, X.~Zhang, R.~Xia, Y.~Song, Y.~Hisai, S.~Aya, M.~Huang, General phase-structure relationship in polar rod-shaped liquid crystals: {Importance} of shape anisotropy and dipolar strength, Giant 11 (2022) 100109.
\newblock \href {https://doi.org/10.1016/j.giant.2022.100109} {\path{doi:10.1016/j.giant.2022.100109}}.

\bibitem{goodby_volume}
Goodby, Mandle, Davis, Zhong, Cowling, What makes a liquid crystal? the effect of free volume on soft matter, Liquid Crystals 42~(5) (2015) 593–622.
\newblock \href {https://doi.org/10.1080/02678292.2015.1030348} {\path{doi:10.1080/02678292.2015.1030348}}.

\bibitem{zannoni_book}
C.~Zannoni, Liquid Crystals and their Computer Simulations, Cambridge University Press, 2022.
\newblock \href {https://doi.org/10.1017/9781108539630} {\path{doi:10.1017/9781108539630}}.

\bibitem{lateral_chain_effect}
E.~Cruickshank, R.~Walker, J.~M.~D. Storey, C.~T. Imrie, The effect of a lateral alkyloxy chain on the ferroelectric nematic phase, RSC Adv. 12 (2022) 29482--29490.
\newblock \href {https://doi.org/10.1039/D2RA05628C} {\path{doi:10.1039/D2RA05628C}}.

\bibitem{gun_shape_nf}
N.~Tufaha, E.~Cruickshank, D.~Pociecha, E.~Gorecka, J.~M. Storey, C.~T. Imrie, Molecular shape, electronic factors, and the ferroelectric nematic phase: Investigating the impact of structural modifications, Chemistry – A European Journal 29~(28) (2023) e202300073.
\newblock \href {https://doi.org/https://doi.org/10.1002/chem.202300073} {\path{doi:https://doi.org/10.1002/chem.202300073}}.

\bibitem{shakespeare_nf}
E.~Cruickshank, P.~Rybak, M.~M. Majewska, S.~Ramsay, C.~Wang, C.~Zhu, R.~Walker, J.~M.~D. Storey, C.~T. Imrie, E.~Gorecka, D.~Pociecha, To be or not to be polar: The ferroelectric and antiferroelectric nematic phases, ACS Omega 8~(39) (2023) 36562--36568.
\newblock \href {https://doi.org/10.1021/acsomega.3c05884} {\path{doi:10.1021/acsomega.3c05884}}.

\bibitem{sulfur_nf}
G.~Stepanafas, E.~Cruickshank, S.~Brown, M.~M. Majewska, D.~Pociecha, E.~Gorecka, J.~M. Storey, C.~T. Imrie, Ferroelectric nematogens containing a methylthio group, Mater. Adv. (2024) 525--538\href {https://doi.org/10.1039/D3MA00446E} {\path{doi:10.1039/D3MA00446E}}.

\end{thebibliography}

\end{document}


\begin{frontmatter}



\title{SUPPLEMENTARY MATERIAL FOR\\A molecular perspective on the emergence of long-range polar order from an isotropic fluid}


\author[inst1]{Aitor Erkoreka}

\affiliation[inst1]{organization={Department of Physics, Faculty of Science and Technology, University of the Basque Country UPV/EHU},
            city={Bilbao},
            country={Spain}}

\author[inst2]{Nerea Sebastián}      
\author[inst2]{Alenka Mertelj}

\affiliation[inst2]{organization={Jožef Stefan Institute},
            city={Ljubljana},
            country={Slovenia}}

\author[inst1]{Josu Martinez-Perdiguero}

\end{frontmatter}

\renewcommand{\thetable}{S1}
\begin{table}[H]
\centering
\begin{tabular}{|c|c|}
\hline
$a$ (\AA{}) & $20.7$  \\ \hline
$b$ (\AA{}) & $5.7$   \\ \hline
$c$ (\AA{}) & $6.5$   \\ \hline
$\mu_{\parallel}$ (D)                  & $11.4$  \\ \hline
$\mu_{\perp}$ (D)                  & $0.3$  \\ \hline
Dipole angle ($^{\circ}$)       & $1.5$   \\ \hline
\end{tabular}
\caption{Molecular parameters and properties of UUQU-4-N obtained from quantum chemical calculations: molecular dimensions ($a$, $b$, $c$), dipole moment parallel and perpendicular to long axis ($\mu_{\parallel}$ and $\mu_{\perp}$), and dipole angle with respect to long axis.}
\end{table}

\pagebreak
\renewcommand{\thefigure}{S1}
\begin{figure}[htbp]
\begin{center}
\includegraphics[width=0.7\textwidth]{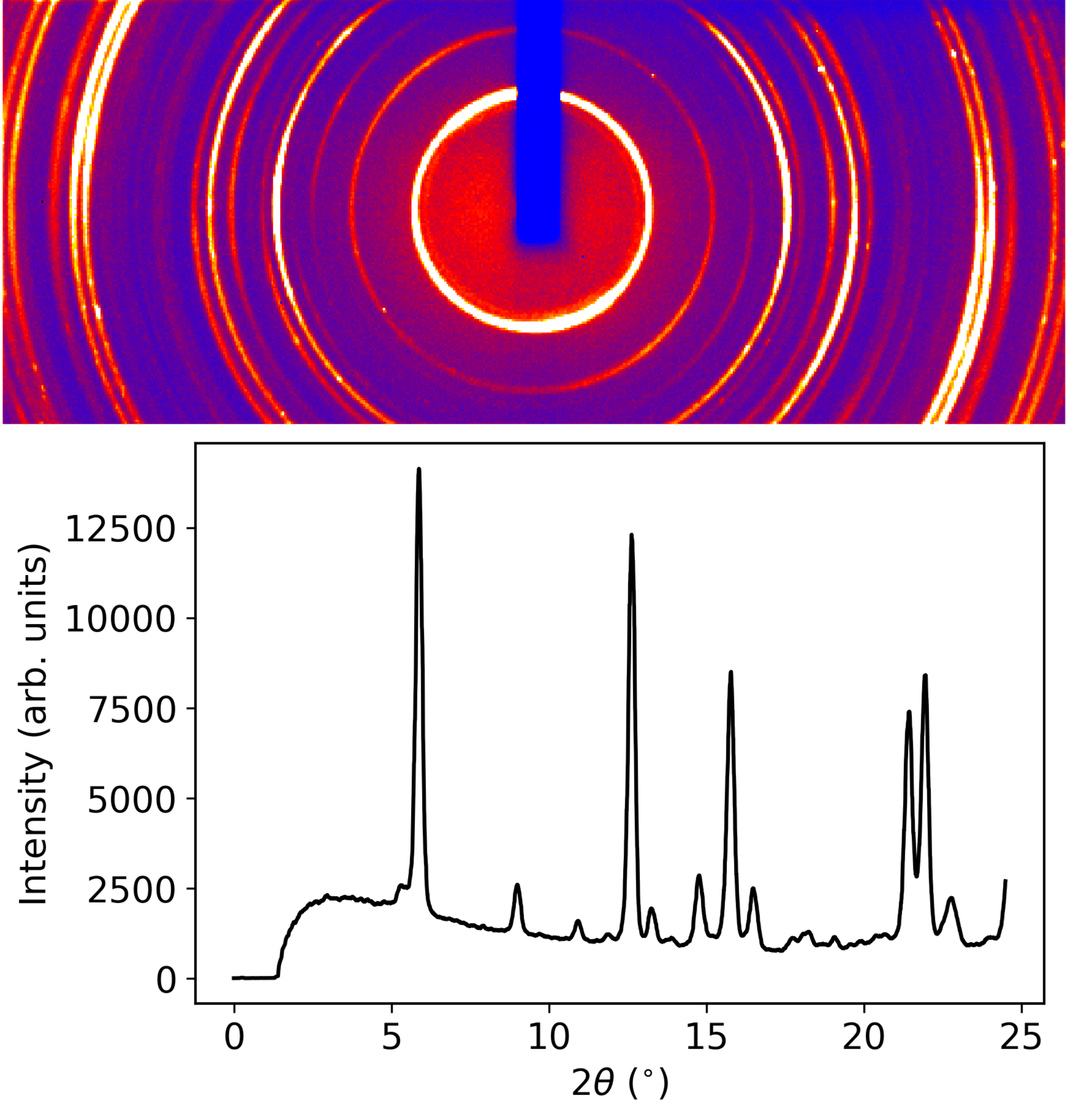}
\end{center}
\caption{\label{fig:xrd_crystal} Top: 2D X-ray diffractogram of UUQU-4-N in the crystalline phase ($1^{\circ}$C). Bottom: the integrated 1D pattern.}
\end{figure}

\pagebreak
\renewcommand{\thefigure}{S2}
\begin{figure}[htbp]
\begin{center}
\includegraphics[width=1.0\textwidth]{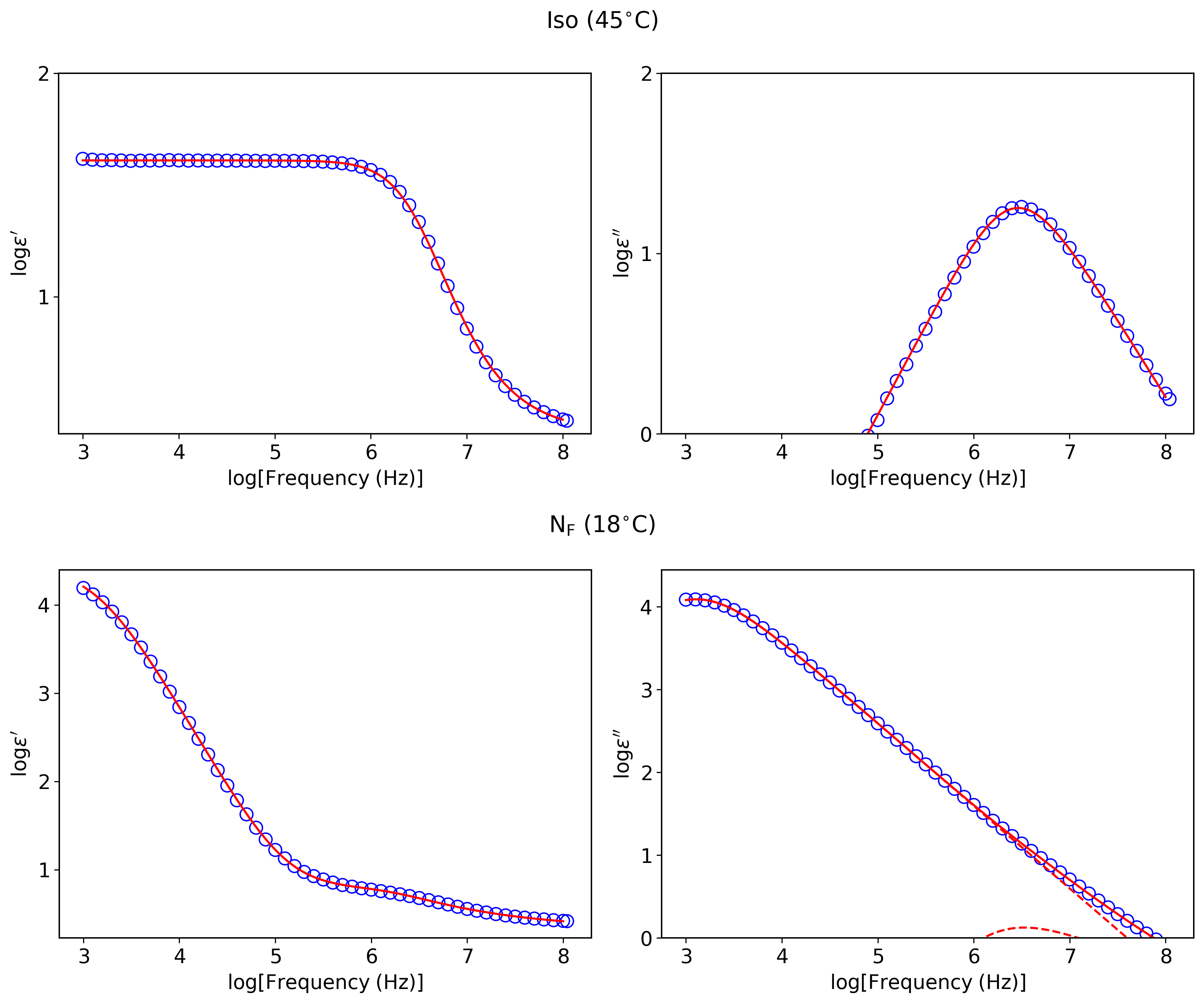}
\end{center}
\caption{\label{fig:fit_examples} Examples of the performed fits according to Eq. (1) in the Iso ($45^{\circ}$C) and N$_{\text{F}}$ ($18^{\circ}$C) phases. The solid red lines correspond to the whole fit, while the dashed red lines correspond to the different mode contributions.}
\end{figure}

\pagebreak
\renewcommand{\thetable}{S2}
\begin{table}[H]
\centering
\begin{tabular}{|c|c|c|}
\hline
Mode              & $\alpha$ & $\beta$      \\ \hline
m$_{\text{Iso}}$  & $1$      & $0.8$--$0.9$ \\ \hline
m$_{\text{NF,L}}$ & $0.9$    & $1.1$        \\ \hline
m$_{\text{NF,H}}$ & $1$      & $0.3$--$0.5$ \\ \hline
\end{tabular}
\caption{$\alpha$ and $\beta$ parameters obtained by fitting the experimental data to Eq. (1). The parameters were let free ($\alpha \leq 1$ and $\alpha \beta \leq 1$) in the fitting procedure so that there are (small) fluctuations around the reported values. The amplitudes and absorption frequencies are plotted in Fig. 4 of the main text.}
\end{table}

\renewcommand{\thefigure}{S3}
\begin{figure}[htbp]
\begin{center}
\includegraphics[width=0.7\textwidth]{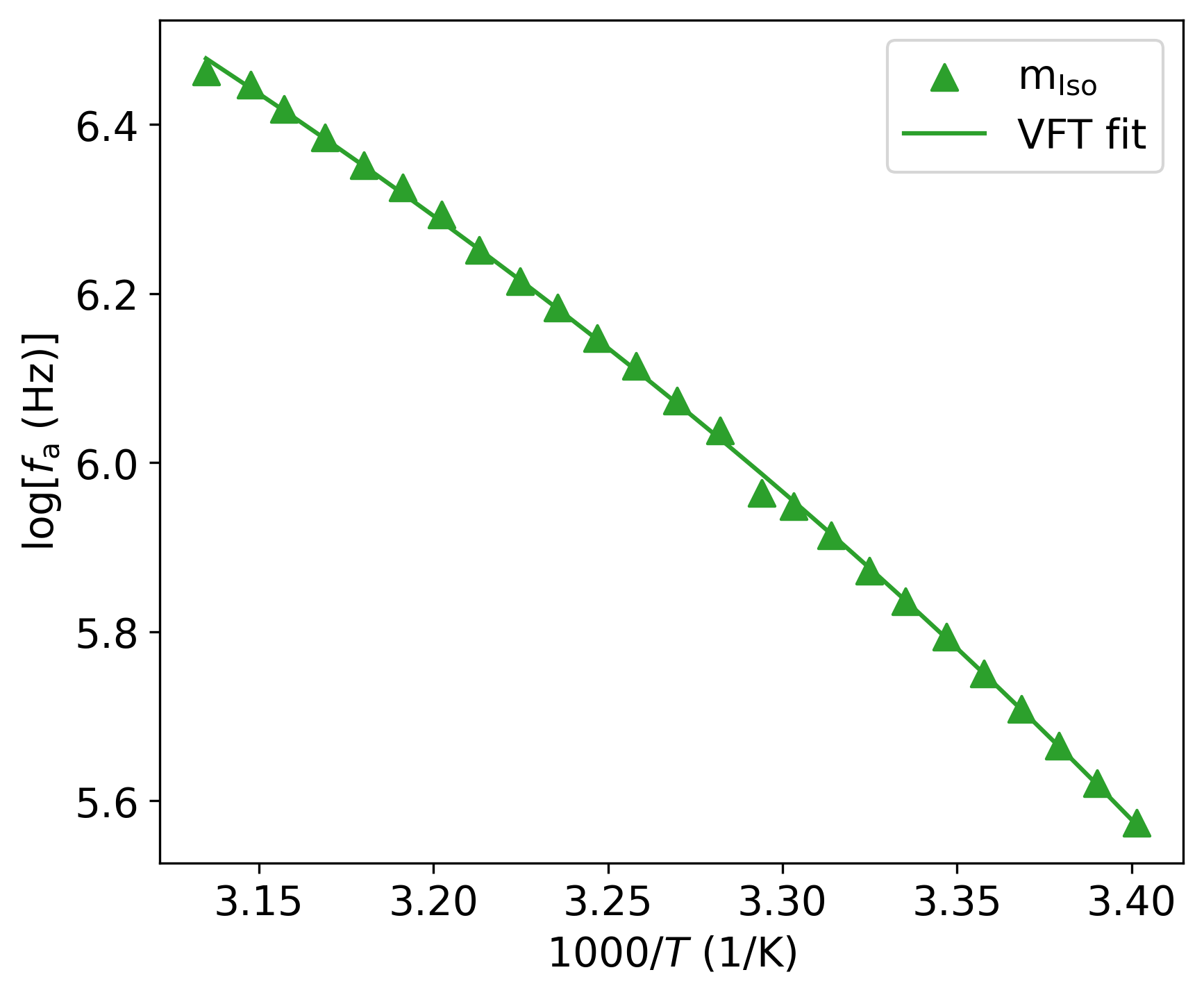}
\end{center}
\caption{\label{fig:vft} Temperature dependence of the absorption frequency of m$_{\text{Iso}}$ and fit to the Vogel-Fulcher-Tammann equation. The fitted parameters are $f_{\infty}=10^9$ Hz, $A=551$ K and $T_0=224$ K.}
\end{figure}

\renewcommand{\thefigure}{S4}
\begin{figure}[htbp]
\begin{center}
\includegraphics[width=0.7\textwidth]{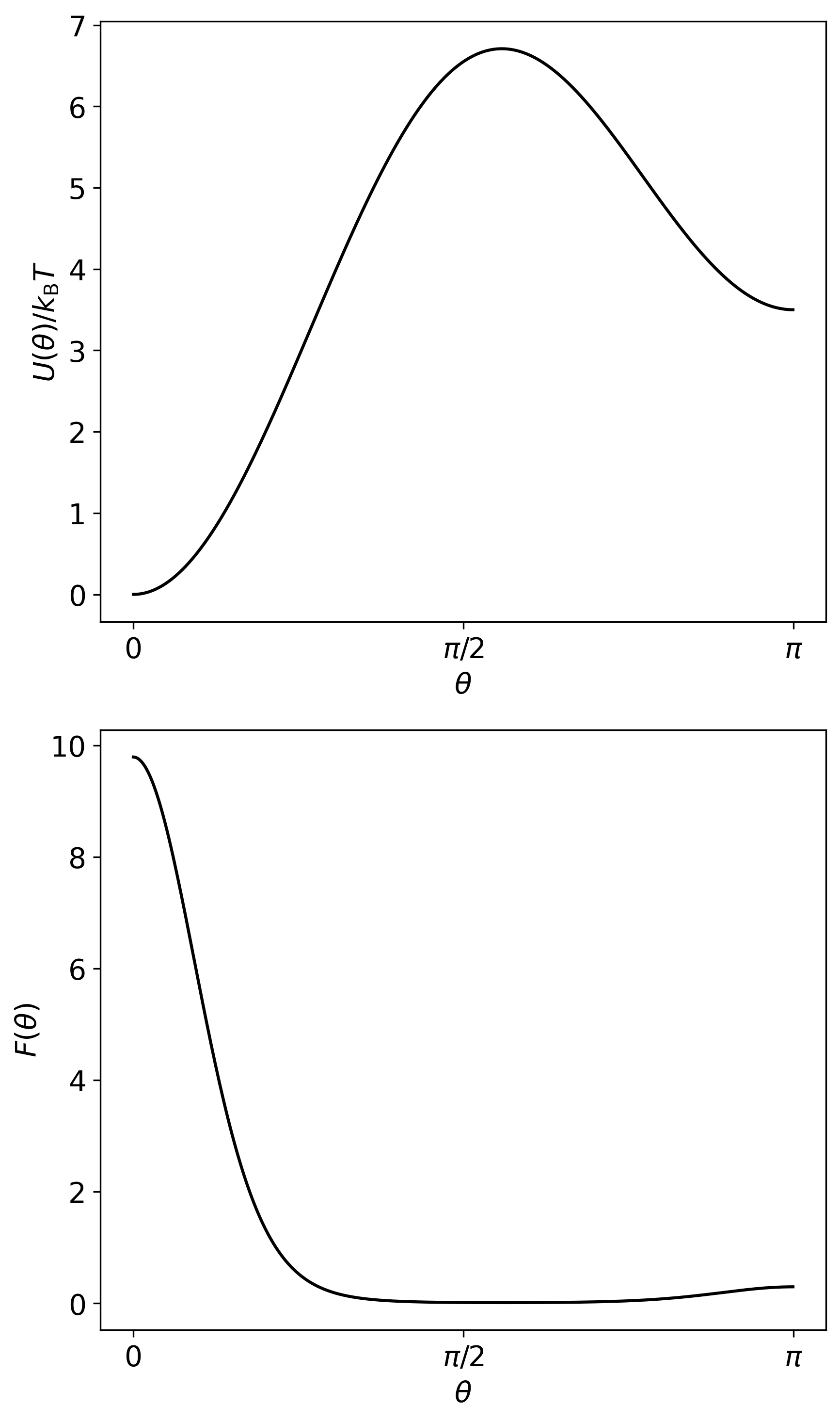}
\end{center}
\caption{\label{fig:shg_model} Nematic potential and normalized distribution function at $18^{\circ}$C in the N$_{\text{F}}$ phase.}
\end{figure}

\newpage
\section*{Hyperpolarizability components}
The relationships between the thermally averaged hyperpolarizability components in the laboratory frame and the molecular ones were obtained assuming that the system is uniaxial, so that the orientational distribution function depends only on $\theta$, i.e. $F=F(\theta)$, and can be written as an infinite sum of Legendre polynomials ($P_n(\cos{\theta})$). When performing the average on the hyperpolarizability tensor only $P_1$ and $P_3$ remain due to symmetry. The expressions are as follows, where the brackets denote thermal averaging:

\begin{equation}
\begin{split}
    \langle \beta_{ZZZ} \rangle=\frac{1}{5}[\left(\langle P_1 \rangle - \langle P_3 \rangle \right)\left(2\beta_{xxz}+2\beta_{yyz}+\beta_{zxx}+\beta_{zyy} \right)\\
    +\left(3\langle P_1 \rangle+2\langle P_3 \rangle \right)\beta_{zzz}].
\end{split}\tag{S1}
\end{equation}

\begin{equation}
\begin{split}
    \langle \beta_{ZXX} \rangle=\frac{1}{10}[\langle P_3 \rangle \left(2\beta_{xxz}+2\beta_{yyz}+\beta_{zxx}+\beta_{zyy}-2\beta_{zzz} \right)\\
    -2\langle P_1 \rangle \left( \beta_{xxz}+\beta_{yyz}-2\beta_{zxx}-2\beta_{zyy}-\beta_{zzz} \right)].
\end{split}\tag{S2}
\end{equation}

\begin{equation}
\begin{split}
    \langle \beta_{XZX} \rangle=\frac{1}{10}[\langle P_3 \rangle \left(2\beta_{xxz}+2\beta_{yyz}+\beta_{zxx}+\beta_{zyy}-2\beta_{zzz} \right)\\
    +\langle P_1 \rangle \left( 3\beta_{xxz}+3\beta_{yyz}-\beta_{zxx}-\beta_{zyy}+2\beta_{zzz} \right)].
\end{split}\tag{S3}
\end{equation}

In the Kleinman approximation, neglecting dispersion effects, $\langle \beta_{ZXX} \rangle=\langle \beta_{XZX} \rangle$.

\renewcommand{\thetable}{S3}
\begin{table}[H]
\centering
\begin{tabular}{|c|c|}
\hline
$\beta_{ijk}$ & ($\times 10^{-30}$ esu) \\ \hline
$\beta_{xxx}$ & $0.0$                   \\ \hline
$\beta_{yxx}$ & $0.0$                   \\ \hline
$\beta_{zxx}$ & $0.1$                   \\ \hline
$\beta_{xyx}$ & $0.0$                   \\ \hline
$\beta_{yyx}$ & $0.0$                   \\ \hline
$\beta_{zyx}$ & $0.1$                   \\ \hline
$\beta_{xyy}$ & $0.0$                   \\ \hline
$\beta_{yyy}$ & $-0.4$                  \\ \hline
$\beta_{zyy}$ & $0.5$                   \\ \hline
$\beta_{xzx}$ & $0.2$                   \\ \hline
$\beta_{yzx}$ & $0.1$                   \\ \hline
$\beta_{zzx}$ & $0.2$                   \\ \hline
$\beta_{xzy}$ & $0.1$                   \\ \hline
$\beta_{yzy}$ & $0.5$                   \\ \hline
$\beta_{zzy}$ & $0.8$                   \\ \hline
$\beta_{xzz}$ & $0.2$                   \\ \hline
$\beta_{yzz}$ & $0.8$                   \\ \hline
$\beta_{zzz}$ & $-2.2$                  \\ \hline
\end{tabular}
\caption{Calculated hyperpolarizability components of the 4-(Difluoromethoxy)-2,6-difluorobenzonitrile segment in the molecular frame, where $z$ is the dipole axis.}
\end{table}

\newpage
\section*{Cluster calculations}
We will only explain the basics of the calculations, while the full details can be found in the original article [27]. Firstly, it should be noted that if rods 1 and 2 are, say, antiparallel, and a third rod is introduced, it is going to be frustrated. Therefore, with respect to a pair in a minimized energy structure with some $\zeta_{21}$ (previously just called $\zeta$), the third rod will take some $\zeta$-shift ($\zeta_{31}$, while $\zeta_{32}=\zeta_{31}-\zeta_{21}$) so as to minimize the total energy of the triplet. The higher the number of neighbors, the closer it will be to a bulk ferroelectric nematic, and the overall structure will change. The following assumptions will be made: the long axes of all the rods are perfectly aligned along the $z$ axis, i.e. $S=\langle P_2(\cos{\theta}) \rangle=1$; the circular cross sections of the rods are arranged in a hexagonal lattice in the $XY$ plane (they do not form layers because of the relative shifts) and rods with maximal overlap with neighbors are said to be in a pseudolayer (see Fig. S5); the distance between the long axes of nearest neighbors is fixed at some value $R_0$, which is a measure of the temperature; since the inter-rod interaction energy falls-off rapidly with $R_0$ (see Fig. S6), we limit the calculation to two pseudolayers with 19 rods each; and, finally, the positive end of the upper rods lie just above the negative end of the lower rods and vice versa. The ferroelectric structure corresponds to all rods being parallel to each other. If rods $1$ and $2$ are antiparallel, there are two possibilities: if rods $3$ and $3^{*}$ in Fig. S5 are parallel, an extended medium will have a net polarization that is $1/3$ of the polarization exhibited by the ferroelectric phase, and corresponds to a ferrielectric structure; however, if rods $3$ and $3^{*}$ in successive rows have opposite orientations, an extended medium will have $0$ polarization, which will be designated as an antiferroelectric structure. From simple geometrical considerations (see Fig. 11 in Ref. [27]), $\zeta_{3*1}=\zeta_{32}$ and $\zeta_{3*2}=\zeta_{32}+\zeta_{21}=\zeta_{31}$. The total energy of the $38$ rods, which consists of $703$ inter-rod interactions, has to be minimized with respect to $\zeta_{21}$ and $\zeta_{31}$, so it is a computationally costly calculation. The ferrielectric structure was found to have a higher energy than the ferroelectric and antiferroelectric structures, so only these two configurations are compared in the text. The charge distribution correspoding to the described case C was assumed, since this was the most favorable one for the emergence of the N$_{\text{F}}$ phase. The results of the calculations are shown in Fig. 10. The graph represents the energy difference between the ferroelectric and antiferroelectric structure, $\Delta E = E_{\text{F}}-E_{\text{AF}}$, as a function of $R_0$. We obtained $\zeta_{21}\approx 4.95$ \AA{} and $\zeta_{31}\approx 9.65$ \AA{} in the ferroelectric case, and $\zeta_{21}\approx 1.20$ \AA{} and $\zeta_{31}\approx 6.25$ \AA{} in the antiferroelectric case.

\renewcommand{\thefigure}{S5}
\begin{figure}[h!]
\begin{center}
\includegraphics[width=0.7\textwidth]{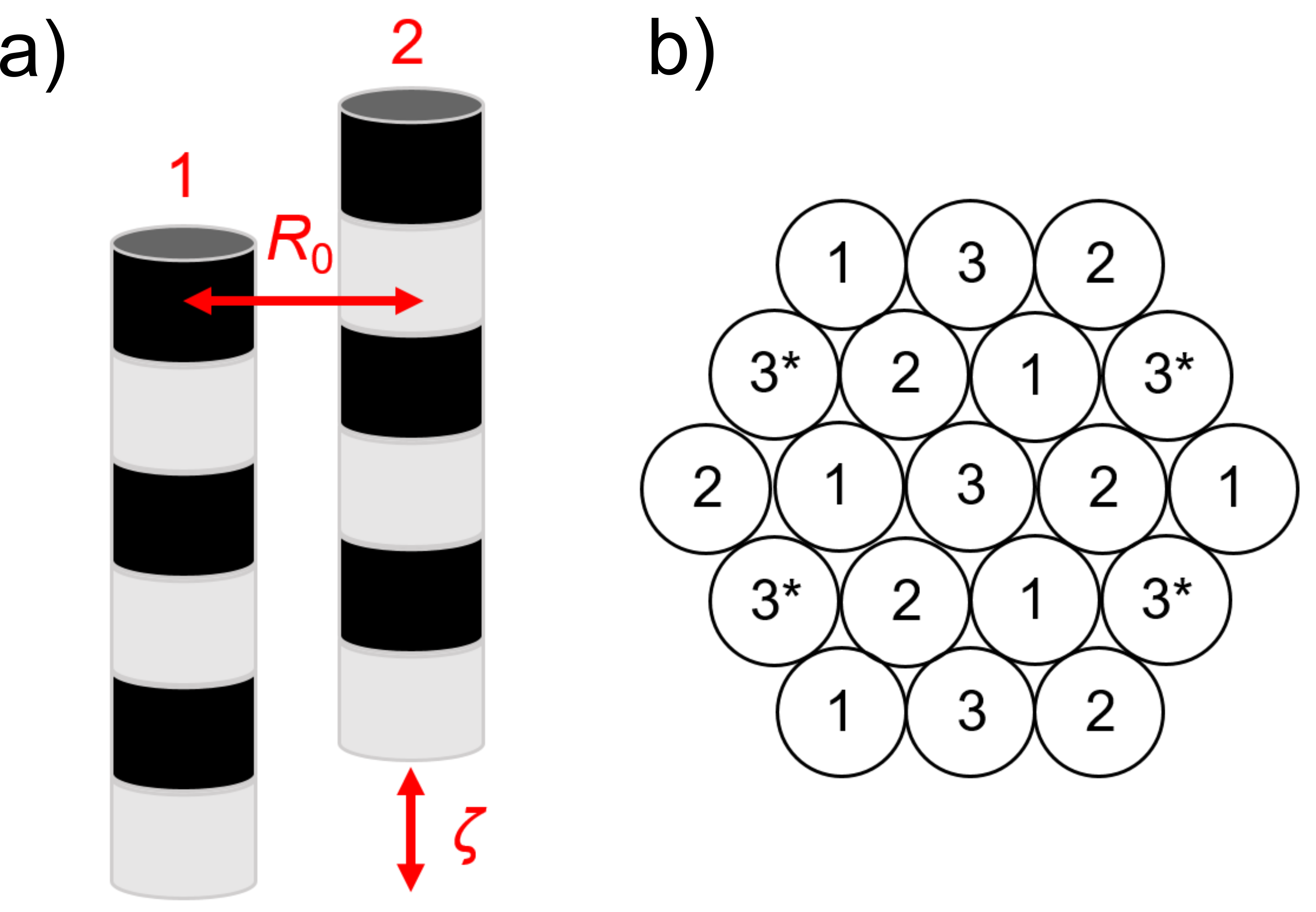}
\end{center}
\caption{\label{fig:lattice} (a) Scheme of the geometry for the calculation of the interaction between two rods with relative shift $\zeta$  (showing the parallel case). (b) Assumed hexagonal arrangement of the circular cross sections of rods of type 1, 2 and 3 in a layer of the cluster as explained in the main article.}
\end{figure}

\renewcommand{\thefigure}{S6}
\begin{figure}[h!]
\begin{center}
\includegraphics[width=0.7\textwidth]{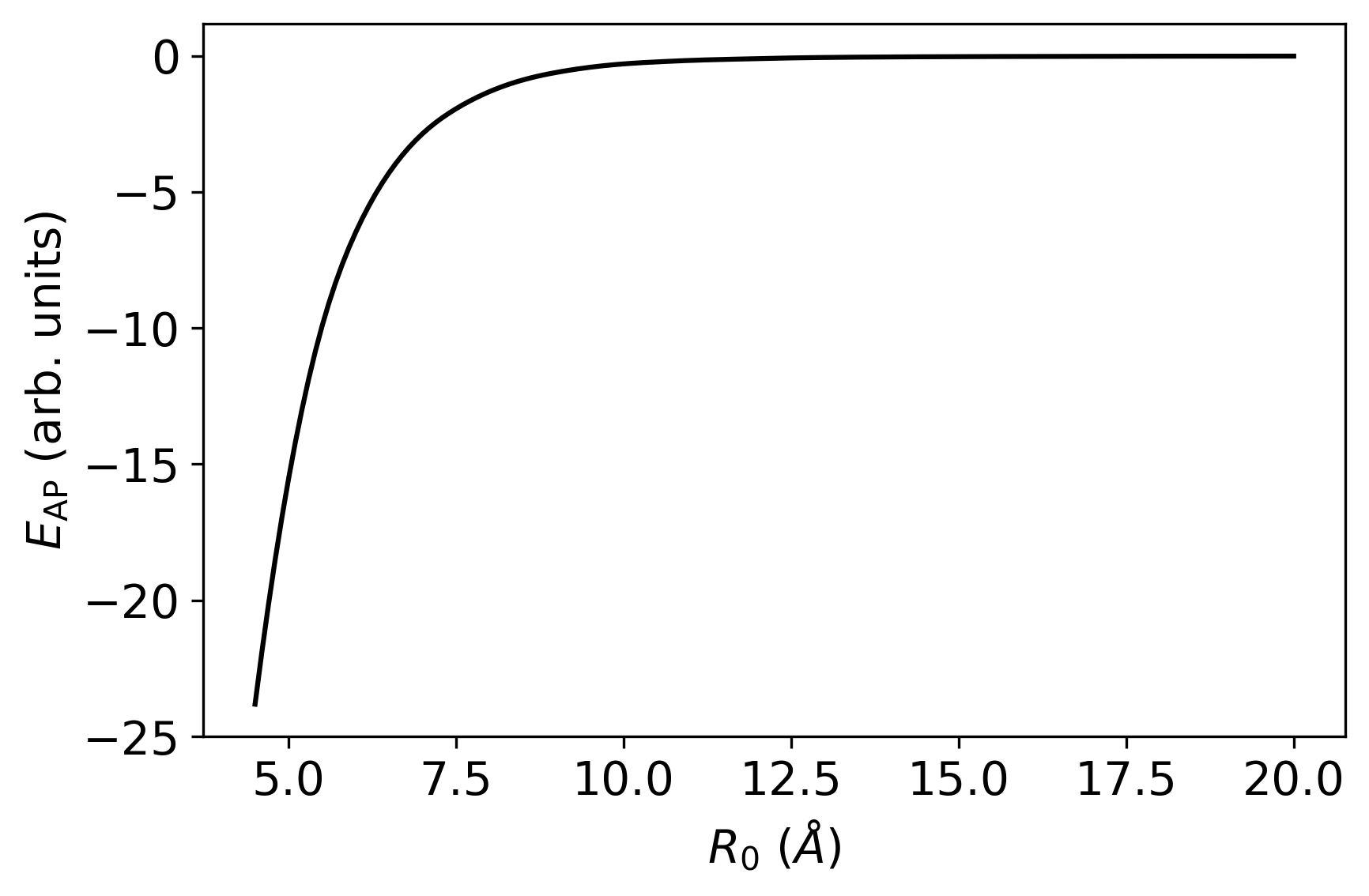}
\end{center}
\caption{\label{fig:model_si} Variation of the electrostatic interaction energy between two rods in the antiparallel configuration with $\zeta_{21}=0$ as a function of $R_0$.}
\end{figure}




